\documentclass{emulateapj}
\usepackage{natbib,aas_macros}
\usepackage{multirow,color}

\begin{document}
\shorttitle{Stellar Populations of LAEs at $z \sim 6$ -- $7$}
\shortauthors{Ono et al.}
\slugcomment{Accepted for publication in ApJ}

\title{%
Stellar Populations of Lyman Alpha Emitters at $\lowercase{z} \sim 6$ -- $7$: Constraints on \\
The Escape Fraction of Ionizing Photons from Galaxy Building Blocks\altaffilmark{\ddag}
}
\author{%
Yoshiaki Ono~\altaffilmark{1},
Masami Ouchi~\altaffilmark{2,3},
Kazuhiro Shimasaku~\altaffilmark{1,4},
James Dunlop~\altaffilmark{5,6}, \\
Duncan Farrah~\altaffilmark{7,8}, 
Ross McLure~\altaffilmark{6}, 
and 
Sadanori Okamura~\altaffilmark{1,4}
}

\email{ono \_at\_ astron.s.u-tokyo.ac.jp}

\altaffiltext{1}{%
Department of Astronomy, Graduate School of Science,
The University of Tokyo, Tokyo 113-0033, Japan
}
\altaffiltext{2}{%
Observatories of the Carnegie Institution of Washington,
813 Santa Barbara Street, Pasadena, CA 91101, USA
}
\altaffiltext{3}{%
Carnegie Fellow
}
\altaffiltext{4}{%
Research Center for the Early Universe, Graduate School of Science,
The University of Tokyo, Tokyo 113-0033, Japan
}
\altaffiltext{5}{%
Department of Physics and Astronomy, University of British Columbia, 
6224 Agricultural Road, Vancouver V6T 1Z1, Canada
}
\altaffiltext{6}{%
SUPA Institute for Astronomy, University of Edinburgh, Royal
Observatory, Edinburgh EH9 3HJ, UK
}
\altaffiltext{7}{%
Department of Astronomy, Cornell University, Ithaca, NY 14853
}
\altaffiltext{8}{%
Astronomy Centre, University of Sussex, Falmer, Brighton, UK
}
\altaffiltext{\ddag}{%
Based on data collected at the Subaru Telescope,
which is operated by the National Astronomical Observatory of Japan.}

\begin{abstract}
We investigate the stellar populations of Ly$\alpha$ emitters (LAEs)
at $z=5.7$ and $6.6$ in a $0.65$ deg$^2$ sky of
the Subaru/\textit{XMM-Newton} Deep Survey (SXDS) Field, using deep
images 
taken with Subaru/Suprime-Cam, UKIRT/WFCAM, and Spitzer/IRAC.
We produce stacked multiband images at each redshift from
$165$ ($z=5.7$) and $91$ ($z=6.6$) 
IRAC-undetected
objects,
to derive typical spectral energy distributions (SEDs) of
$z\sim 6$ -- $7$ LAEs for the first time.
The stacked LAEs have as blue UV continua as
the HST/WFC3 $z$-dropout galaxies of similar $M_{\rm UV}$,
with a spectral slope $\beta \sim -3$, 
but at the same time
they have red UV-to-optical colors with detection in the $3.6\mu$m band.
Using SED fitting we find that the stacked LAEs have low stellar
masses of $\sim \left( 3-10 \right) \times 10^{7}M_\odot$, very young ages of
$\sim 1-3$ Myr, negligible dust extinction, and strong
nebular emission from the ionized 
interstellar medium, 
although the $z=6.6$ object
is fitted similarly well with high-mass models without nebular emission;
inclusion of nebular emission reproduces the red UV-to-optical
colors while keeping the UV colors sufficiently blue.
We infer that typical LAEs at $z\sim 6-7$ are building blocks
of galaxies seen at lower redshifts.
We find a tentative decrease in the Ly$\alpha$ escape fraction
from $z=5.7$ to $6.6$, which may imply an increase in the 
intergalactic medium
neutral fraction.
From the minimum contribution of nebular emission required to fit
the observed SEDs, we place an upper limit on the escape fraction
of ionizing photons to be $f_{\rm esc}^{\rm ion} \sim 0.6$ at $z=5.7$
and $\sim 0.9$ at $z=6.6$. 
We also compare the stellar populations of our LAEs with
that of stacked HST/WFC3 $z$-dropout galaxies.
\end{abstract}

\keywords{%
cosmology: observations ---
galaxies: formation ---
galaxies: evolution ---
galaxies: high-redshift ---
galaxies: stellar content ---
}


\section{INTRODUCTION} \label{sec:introduction}

Ly$\alpha$ emitters (LAEs) are a galaxy population commonly seen
at high redshift.
The high number density of LAEs indicates their importance in the
evolutionary history of galaxies.
Since galaxies with strong Ly$\alpha$ emission can be identified
as LAEs even when their continuum emission is too faint to be
detected in broadband imaging, LAEs provide an opportunity to
probe low-mass galaxies with active star-formation.
Some
of them may be building blocks of more evolved galaxies.

Surveys of LAEs have been made primarily with narrow-band imaging
to isolate Ly$\alpha$ emission
\citep[e.g.,][]{hu1998,rhoads2000,iye2006},
and until now over a thousand LAEs have been photometrically
selected and/or spectroscopically identified \citep[e.g.,][]{hu2002,ouchi2003,malhotra2004,taniguchi2005,kashikawa2006,shimasaku2006,dawson2007,gronwall2007,murayama2007,ouchi2008,shioya2009,nilsson2009,guaita2009,hayes2010b}.

Studying the stellar population of LAEs is essential to understand
their physical nature and to reveal the relationship between LAEs
and other high-redshift galaxies
such as Lyman break galaxies \citep[e.g.,][]{steidel1996a}.
At $z \simeq 3$ -- $5$, much progress has been made recently,
and it has been revealed from large samples with multiband
photometry that most LAEs are small galaxies with masses
$10^8$ -- $10^9 M_\odot$ and young ages $<10^8$ yr
\citep[e.g.,][]{gawiser2007,nilsson2007,pirzkal2007},
while some have large stellar masses of $\simeq 10^{10} M_\odot$
\citep[e.g.,][]{lai2008,finkelstein2009,ono2009}.

At higher redshifts, a few studies have reported on stellar
populations of LAEs, but they are all based on a very small sample
and consensus has not been reached.
\cite{lai2007} have studied stellar populations of three bright
LAEs at $z = 5.7$ detected in rest-frame optical wavelengths,
and found that
they have high stellar masses of $\sim 10^{10} M_\odot$,
as old ages as the Universe at their redshifts,
and some amount of dust.
\cite{ouchi2009} have reported the discovery of a giant LAE at
$z=6.595$, Himiko, with a Spitzer/Infrared Array Camera
\citep[IRAC;][]{fazio2004} counterpart, and estimated its stellar
mass to be ($0.9$ -- $5.0$) $\times 10^{10} M_\odot$.
It should be noted that these four LAEs are all detected in at
least one IRAC band, which means that they are not typical LAEs
but rare, massive-end objects among the overall LAE population.
On the other hand, \cite{pirzkal2007} have studied three faint
LAEs at $5.2 \leq z \leq 5.8$ 
found by Hubble Space Telescope (HST)/Advanced Camera for Survey
(ACS) slitless spectroscopy in the Hubble Ultra Deep Field (HUDF),
to show that they are all very young ($\simeq$ several
$\times 10^6$ yr) with low masses ($\simeq 10^6$ -- $10^8 M_\odot$)
and small dust extinctions ($A_V = 0$ -- $0.1$).
\cite{chary2005} have studied a gravitationally-lensed LAE at
$z=6.56$, HCM6A, and reported that it has a stellar mass of
$\sim 10^9 M_\odot$ and very young age ($\sim 5$ Myr).
Although these two studies may be picking up typical LAEs at
each redshift, it is hard to draw robust conclusions from such
small numbers of objects.

Recently, \cite{ouchi2008} and \cite{ouchi2009} have constructed
the largest available sample of $z = 5.7$ and $6.6$ LAEs
in an about $1$ deg$^2$ of the Subaru/\textit{XMM-Newton} Deep Field (SXDF)
from deep optical broadband and narrowband data.
About $65${\%} of the field is also covered by deep $JHK$ images
taken with the United Kingdom Infrared Telescope (UKIRT)/
Wide Field Infrared Camera \citep[WFCAM;][]{casali2007} from
UKIRT Infrared Deep Sky Survey (UKIDSS) Ultra Deep Survey
\citep[UDS;][]{lawrence2007,warren2007},
and $3.6$ -- $8.0\mu$m images taken with the
Spitzer/IRAC from the Spitzer legacy survey of the UDS field
(SpUDS; 
Spitzer Proposal ID {\#}40021;
PI: J. Dunlop). 
In this paper we use these wide-field, multiwaveband survey data
to study the stellar population of typical LAEs at $z=5.7$ and 6.6
from stacking a large number of faint objects (165 at $z=5.7$ and
91 at $z=6.6$).
At each redshift the stacked object is detected in several broadbands
including the $3.6\mu$m band, enabling us to place meaningful
constraints on stellar population parameters.
We also examine the stellar population of $z$-dropout galaxies
recently discovered in the deep HST/Wide Field Camera 3 (WFC3)
data of the HUDF, 
using the stacked spectral energy distribution (SED) 
obtained by \cite{labbe2010} from $14$ objects of the \cite{oesch2010} sample.

As noted by \cite{schaerer2009},
careful SED fittings including
nebular emission are required to accurately determine stellar
population parameters for high-$z$ galaxies
\citep[see also][]{zackrisson2008,raiter2010,schaerer2010}.
Nebular emission is produced in the interstellar medium (ISM) ionized by hot stars.
Hence, the nebular emission of a galaxy increases as the fraction
of ionizing photons absorbed by HI gas in the ISM increases,
or equivalently, as the fraction of ionizing photons escaping
from the galaxy, $f_{\rm esc}^{\rm ion}$, decreases.
To see the influence of nebular emission on the determination
of stellar population parameters,
we examine SED models for two extreme $f_{\rm esc}^{\rm ion}$ values:
$f_{\rm esc}^{\rm ion}=1$ where the SED is determined solely by stellar
emission as most previous studies assumed,
and $f_{\rm esc}^{\rm ion}=0$ where
the contribution of nebular emission to the SED is largest.
We find that the latter models generally give a very good fit
to the observed SEDs, especially the $z=5.7$ one which is fit
by the latter significantly better.

In addition,
we treat $f_{\rm esc}^{\rm ion}$ as a free parameter in our SED fitting
to place rough constraints on $f_{\rm esc}^{\rm ion}$, because redshifts
of $z\sim 6$ -- $7$ are close to the end of cosmic reionization \citep[e.g.,][]{fan2006,becker2007}
and the $f_{\rm esc}^{\rm ion}$ of galaxies is a key parameter which
determines the ionizing photon budget.
At redshifts up to $z \sim 4$, the $f_{\rm esc}^{\rm ion}$ of galaxies
has been estimated or constrained by a variety of methods using,
e.g., FUV spectra and narrow-band images of star-forming galaxies
\citep[e.g.,][]{steidel2001,shapley2006,iwata2009}
and the distribution of neutral hydrogen column densities in
the after-glow spectra of long duration GRBs
\citep[e.g.,][]{chen2007,fynbo2009},
although the results do not agree well with each other.
For example, \cite{shapley2006} have derived
$f_{\rm esc}^{\rm ion} \geq 0.65$
for two $z \sim 3$ LBGs with detected Lyman continua
and
$f_{\rm esc}^{\rm ion} = 0.14$ from
a composite spectrum of $14$ $z \sim 3$ LBGs\footnote{They 
have assumed the intrinsic UV to Lyman
continuum flux density ratio to be $3.0$,
and corrected for IGM absorption.},
while \cite{chen2007} have placed an upper limit of
$f_{\rm esc}^{\rm ion} \leq 0.075$ using a
compiled sample of $28$ GRBs \citep[see also][]{fynbo2009}.

At redshifts as high as $z \sim 6$ -- $7$, most constraints
are based on the UV luminosity density of galaxies.
For example, \cite{ouchi2009b} and \cite{finkelstein2009f}
inferred the production rate of ionizing photons in galaxies
from the observed UV luminosity density of $z$-dropout galaxies
and obtained lower limits to $f_{\rm esc}^{\rm ion}$
to keep the intergalactic medium (IGM) ionized.
However, this method has a number of uncertainties such as
the UV luminosity function of galaxies and the clumpiness of
the IGM.
Recently, \cite{bouwens2010b} proposed that the very blue UV color
of $z$-dropout galaxies they found may be due to weak nebular
emission and hence high $f_{\rm esc}^{\rm ion}$, because strong nebular
emission makes the UV color too red.
This is interesting in relating the SED to $f_{\rm esc}^{\rm ion}$.
However, as they already state, $f_{\rm esc}^{\rm ion}$ is not uniquely
determined from the UV color but it also depends on the age of
the stellar population.
In this paper we show that inclusion of a UV-to-optical color
reduces this degeneracy and 
provides constraints on $f_{\rm esc}^{\rm ion}$ for our LAEs.

The outline of this paper is as follows.
After describing the imaging data used in this study in Section 2,
we produce stacked images in Section 3.
Our SED fitting method is presented in Section 4.
In Section 5, we present and discuss our SED fitting results
and constraints on $f_{\rm esc}^{\rm ion}$.
A summary is given in Section 6.
Throughout this paper, we use magnitudes in the AB system
\citep{oke1983} and assume a flat universe with
($\Omega_m$, $\Omega_\Lambda$, $h$) $=$ ($0.3$, $0.7$, $0.7$).
The age of the universe is $\simeq 0.98$ Gyr at $z = 5.70$ and
$\simeq 0.82$ Gyr for $z = 6.56$.

\begin{figure*}./
 \begin{center}
  \includegraphics[scale=1.0]{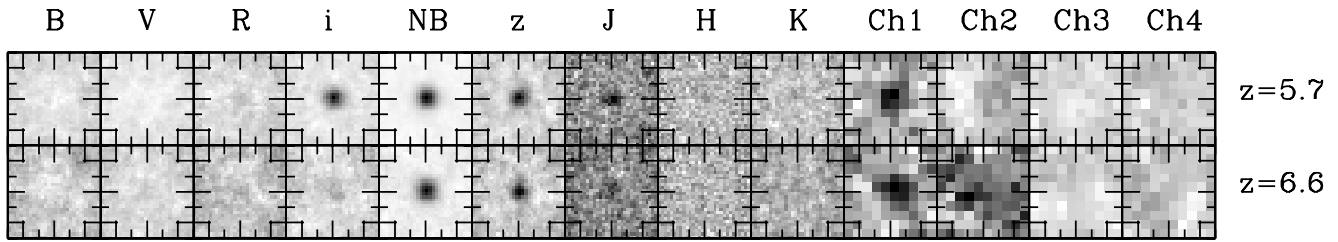}
\end{center}
 \caption[]
{
Median-stacked images of LAEs at $z=5.7$ (top) and $z=6.6$ (bottom)
taken by the Subaru Suprime-Cam ($B$, $V$, $R$, $i'$, $NB$,
$z'$), the UKIRT WFCAM ($J$, $H$, $K$),
and the Spitzer IRAC ($3.6\mu$m, $4.5\mu$m, $5.8\mu$m, $8.0\mu$m).
The $z=5.7$ LAE is detected (brighter than the $3\sigma$ magnitude)
in $i'$, $NB816$, $z'$, and $3.6\mu$m, and nearly detected in $J$.
The $z=6.6$ LAE is detected in $NB921$, $z'$, and $3.6\mu$m.
Each panel is $6'' \times 6''$ in size, or $\simeq 35$ $(33)$ kpc
at $z = 5.70$ $(6.56)$.
}
\label{fig:faces}
\end{figure*}

\begin{deluxetable}{cccccc} 
\tablecolumns{3} 
\tablewidth{0pt} 
\tablecaption{Median-Stacked Lyman Alpha Emitters \\
at $z=5.7$ and $6.6$ \label{tab:photo_prop}}
\tablehead{
\colhead{}    & \colhead{$z=5.7$ LAE}    & \colhead{$z=6.6$ LAE}
}
\startdata 
$NB$\tablenotemark{a}  &  $25.5$ ($27.6$)  &  $25.6$ ($27.0$) \\
$m_{\rm con}$\tablenotemark{b}  &  $27.2$ ($28.2$)  &  $27.4$ ($26.6$) \\
$B$  &  $99.9$ ($30.4$)  &  $99.9$ ($30.1$) \\
$V$  &  $99.9$ ($29.8$)  &  $99.9$ ($29.5$) \\
$R$  &  $99.9$ ($29.9$)  &  $99.9$ ($29.7$) \\
$i'$  &  $27.9$ ($30.0$)  &  $29.9$ ($29.6$) \\
$z'$  &  $27.5$ ($29.0$)  &  $27.7$ ($28.6$) \\
$J$  &  $27.6$ ($27.5$)  &  $27.7$ ($27.1$) \\
$H$  &  $99.9$ ($26.6$)  &  $99.9$ ($26.3$) \\
$K$  &  $99.9$ ($27.0$)  &  $99.9$ ($26.7$) \\
$3.6 \mu$m  &  $26.7$ ($27.0$)  &  $26.6$ ($26.7$) \\
$4.5 \mu$m  &  $32.0$ ($26.6$)  &  $26.7$ ($26.3$) \\
$5.8 \mu$m  &  $99.9$ ($25.4$)  &  $99.9$ ($24.4$) \\
$8.0 \mu$m  &  $99.9$ ($25.3$)  &  $99.9$ ($24.1$) \\
$\beta$\tablenotemark{c}  &  $-2.9 \pm 1.0$  &  $-3.0 \pm 2.7$ \\
$f({\rm Ly}\alpha)$ [$10^{-18} $erg s$^{-1}$cm$^{-2}$]\tablenotemark{d}  &  $11.0 \pm 0.77$  &  $8.93 \pm 1.25$ \\
$L({\rm Ly}\alpha)$ [$10^{42} $erg s$^{-1}$]\tablenotemark{d}  &  $3.90 \pm 0.27$  &  $4.37 \pm 0.61$ \\
${\rm EW}({\rm Ly}\alpha)_{\rm rest}$ [{\AA}]\tablenotemark{d}  &  $78.1^{+96.3}_{-65.4}$  &  $84.2^{+170.8}_{-49.3}$ \\
redshift\tablenotemark{e} & $5.70$ & $6.56$ 
\enddata 
\tablecomments{
All magnitudes are total magnitudes.
$99.9$ mag means negative flux densities.
Magnitudes in parentheses are $3\sigma$ limiting magnitudes.
}
\tablenotetext{a}{%
$NB816$ for $z=5.7$ LAE, $NB921$ for $z=6.6$ LAE.
}
\tablenotetext{b}{%
Rest-frame UV continuum magnitudes at (observed) effective
wavelengths of $8303${\AA} ($z=5.7$) and $9445${\AA} ($z=6.6$)
after correction for Ly$\alpha$ emission and the IGM absorption,
derived from $NB816$ and $i'$ magnitudes ($z=5.7$) and $NB921$
and $z'$ magnitudes ($z=6.6$).
}
\tablenotetext{c}{%
Derived from $m_{\rm con}$ and $J$ magnitude using eq. (\ref{eq:beta})
}
\tablenotetext{d}{%
Derived assuming that the redshifted Ly$\alpha$ line is
at the center of the $NB$ filter.
}
\tablenotetext{e}{%
Corresponding to the central wavelengths of $NB816$ and $NB921$.
}
\end{deluxetable} 

\section{DATA} \label{sec:data}

Deep $BVRi'z'$ images of the SXDF were taken with Suprime-Cam
\citep{miyazaki2002} on the Subaru Telescope by the
Subaru/\textit{XMM-Newton} Deep Survey project
\citep[SXDS;][]{furusawa2008}.
\cite{ouchi2008} and \cite{ouchi2009} combined this public data set
with their own imaging data taken with Suprime-Cam
through two narrowband filters,
$NB816$ ($\lambda_c = 8150${\AA}, FWHM $= 120${\AA})
and
$NB921$ ($\lambda_c = 9196${\AA}, FWHM $= 132${\AA}),
and constructed samples of $401$ $z=5.7$ and $207$ $z=6.6$ LAEs
over a sky area of $\simeq 1$ deg$^2$  
by applying the following selection criteria:
(i) existence of a narrowband excess, 
(ii) no detection in any bandpasses blueward of the Lyman limit,
and (iii) existence of a spectral break due to IGM absorption. 
The $z=5.7$ LAEs have 
$L$(Ly$\alpha$) $\gtrsim 3 \times 10^{42}$ erg s$^{-1}$ 
and 
EW(Ly$\alpha$) $\gtrsim 27$ {\AA} 
\citep{ouchi2008}, 
and the $z=6.6$ LAEs have 
$L$(Ly$\alpha$) $\gtrsim 3 \times 10^{42}$ erg s$^{-1}$ 
and 
EW(Ly$\alpha$) $\gtrsim 14$ {\AA} 
\citep{ouchi2010}.

About $77${\%} of the SXDF Suprime-Cam field was imaged
in the $J$, $H$, and $K$ bands
with the wide-field near infrared camera WFCAM on the UKIRT
in the UKIDSS/UDS project \citep{lawrence2007}.
The UKIDSS/UDS is underway, and we use Data Release 5 for this study.
We align the $J,H,K$ images with the SXDS optical images
using common, bright stars, and then smooth them
with Gaussian filters so that the PSF sizes of the $J,H,K$ images
match those of the optical images (FWHM $\approx 1.''0$).
The $3\sigma$ limiting magnitudes over a $2''$-diameter aperture
are calculated to be $24.5$, $24.2$, and $24.4$ in the $J,H,K$ bands,
respectively.
Because the zero-point magnitudes for the $J,H,K$ images are given
in the Vega system, we convert them into AB magnitudes using
the offset values given in Table 7 of \cite{hewett2006}.

The SpUDS covers $0.65$ deg$^2$ of the overlapping area of the
SXDS and UDS fields.
This $0.65$ deg$^2$ area corresponds to an effective survey volume of
$6.0 \times 10^5$ Mpc$^3$ for $z=5.7$ LAEs and
$5.2 \times 10^5$ Mpc$^3$ for $z=6.6$ LAEs, respectively.
All of the SpUDS IRAC images are geometrically matched to the
optical images.
We calculate the $3\sigma$ limiting magnitude over
a $3''$-diameter aperture to be
$24.8$, $24.5$, $22.7$, and $22.6$ in the $3.6$, $4.5$, $5.8$,
and $8.0 \mu$m IR AC bands, respectively.

\section{STACKING ANALYSIS} \label{sec:stacking_analysis}

In this paper, we only analyze LAEs in the overlapping area of
$0.65$ deg$^2$ where the Suprime-Cam, WFCAM, and IRAC data are
all available.
We perform IRAC Channel 1 photometry with a $3''$-diameter aperture
at the position of LAEs in the narrowband images, using the IRAF
task \verb|phot|.
Among a total of $189$ ($106$) LAEs at $z=5.7$ ($6.6$),
$165$ ($91$) are found to be fainter than
the IRAC $3.6\mu$m-band $3 \sigma$ magnitude (i.e., $24.8$ mag),
which means that
they are neither rare massive-end objects among the overall LAE population, 
nor confused by neighboring objects.
We make their median-stacked multi-waveband images separately
for the two redshifts.
Their cutouts are shown in Figure \ref{fig:faces}.

Among those not used for stacking,
some seem to have counterparts in the IRAC $3.6\mu$m image.
However, they have not been spectroscopically confirmed,
except for the giant LAE Himiko already studied by \cite{ouchi2009}.
We will discuss elsewhere stellar populations of these bright LAE
candidates after we confirm their redshifts by spectroscopy.

\subsection{Photometry} \label{subsec:photometry}

We perform $BVRi'z'JHK$ photometry with a $2''$-diameter aperture
at the position of the LAEs in the narrowband images,
using the IRAF task \verb|phot|.
We then convert them into total magnitudes
by subtracting aperture correction terms\footnote{Aperture correction terms [ABmag] are
$0.147 (B)$, $0.115 (V)$, $0.146 (R)$, $0.202 (i')$, $0.156 (z')$,
$0.196 (J)$, $0.208 (H)$, $0.173 (K)$.
},
which are evaluated for bright and isolated point sources in each band.
To evaluate the aperture correction term for the two narrow bands,
we measure fluxes for bright point sources in a series of apertures
from $2''$ up to $6''$ with an interval of $0.1''$.
Since we find that the fluxes level off for $> 5''$ apertures,
we define the difference in magnitude between the $2''$ and $5''$
aperture magnitudes as the aperture correction term.
For the Spitzer/IRAC four bands, we measure $3''$-diameter aperture
magnitudes for each LAE and converted them to total magnitudes by
applying the aperture correction
given by Multiwavelength Survey by Yale-Chile (MUSYC)
survey\footnote{\texttt{http://data.spitzer.caltech.edu/popular/simple/
}\\\texttt{20070601{\_}enhanced/doc/00README{\_}photometry}}.
The correction values are $0.52$, $0.55$, $0.74$, and $0.86$ mag
for $3.6\mu$m, $4.5\mu$m, $5.8\mu$m, and $8.0\mu$m, respectively.
We measure the limiting magnitude for each band
by making $1000$ median-stacked sky noise images,
each of which is made of ($165$ for $z=5.7$ and $91$ for $z=6.6$)
randomly-selected sky noise images.
Table \ref{tab:photo_prop} summarizes their photometry,
UV spectral slope $\beta$ (see Section \ref{subsec:UV_spectral_slope}),
average $f({\rm Ly}\alpha)$, $L({\rm Ly}\alpha)$,
and EW$({\rm Ly}\alpha)$.
We find that the both stacked objects are brighter than
the $3 \sigma$ magnitude in the $3.6\mu$m band.

Since our $i'$-band photometry for the $z=5.7$ LAE
and $z'$-band photometry for the $z=6.6$ LAE
are contaminated by Ly$\alpha$ emission and 
IGM absorption, 
we derive the emission-free continuum magnitude $m_{\rm con}$
at $\simeq 8303${\AA} for $z=5.7$
and at $\simeq 9445${\AA} for $z=6.6$, respectively.
We obtain $m_{\rm con} = 27.22 \pm 0.14$ from $NB816$- and
$i'$-band photometry,
and $m_{\rm con} = 27.43 \pm 0.57$ from $NB921$- and $z'$-band
photometry,
taking account of the contributions of Ly$\alpha$ emission
and IGM absorption \citep{madau1995} to the photometry in
each bandpass \citep{shimasaku2006}.
We use $m_{\rm con}$ instead of the $i'$-band ($z'$-band) magnitude 
for the $z=5.7$ ($6.6$) LAE
to derive the UV spectral slope $\beta$ in
Section \ref{subsec:UV_spectral_slope}
and constrain the stellar population in Section \ref{sec:sedfitting}.

\subsection{UV Spectral Slope}\label{subsec:UV_spectral_slope}

\begin{figure}
 \begin{center}
  \includegraphics[scale=0.6]{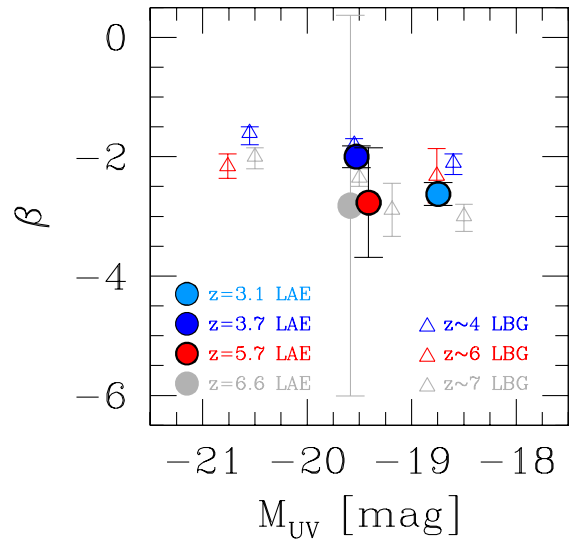}
\end{center}
 \caption[]
{
UV continuum slope $\beta$ versus rest-frame UV absolute magnitude
$M_{\rm UV}$.
The filled circles and open triangles represent LAEs and LBGs,
respectively, colored according to their redshifts.
The red and gray circles are our LAEs at $z=5.7$ and $6.6$,
respectively,
where the $M_{\rm UV}$ of the $z=6.6$ LAE has been
shifted $-0.15$ for clarity.
Also shown are the values for stacked LAEs at $z=3.1$ (cyan circle)
and $z=3.7$ (blue circle) \citep[]{ono2009},
$z \sim 4$ LBGs \citep[blue triangles:][]{bouwens2010b},
$z \sim 6$ LBGs \citep[red triangles:][]{bouwens2009b},
and $z \sim 7$ LBGs \citep[gray triangles:][]{bouwens2010b,labbe2010}.
}
\label{fig:betas}
\end{figure}

We estimate the slope of the rest-frame UV continuum, $\beta$,
from two broad-band magnitudes, $m_1$ and $m_2$:
\begin{equation}
\beta
       =-\frac{m_1 - m_2}{2.5 \log \left( \lambda_{\rm c}^1 / \lambda_{\rm c}^2 \right)} - 2.
\label{eq:beta}
\end{equation}
where $\lambda_{\rm c}^1$ and $\lambda_{\rm c}^2$
are the central wavelengths of the two broadbands.
For the $z=5.7$ LAE,
we set $(m_1, m_2) = (m_{\rm con},m_{\rm J})$
and obtain $\beta = - 2.9 \pm 1.0$,
where $m_{\rm con}$ is the continuum magnitude at
8303 \AA\ (see Table 1) and
$\lambda_{\rm c}^{\rm J}=12500$ \AA\ \citep{tokunaga2002}. 
These wavelengths correspond to rest-frame 
$1240$ {\AA} and $1870$ {\AA}, respectively. 
Similarly, $\beta = - 3.0 \pm 2.7$ is obtained for the $z=6.6$ LAE
from the continuum magnitude at 9445 {\AA} and $m_{\rm J}$, 
corresponding to $1250$ {\AA} and $1650$ {\AA}, respectively.
Thus both have a very blue slope, although the uncertainty
is large especially for the $z=6.6$ object.

Figure \ref{fig:betas} plots $\beta$ against the rest-frame UV
absolute magnitude, $M_{\rm UV}$, for our objects together with
LAEs at two lower redshifts ($z=3.1$ and $3.7$) and LBGs at
$z \sim 4$, $6$, and $7$.
For the LAEs we calculate $\beta$ using our own data \citep{ono2009}:
$\beta = -2.6 \pm 0.2$ for $z=3.1$ and $\beta = -2.0 \pm 0.2$
for $z=3.7$ from $R$ and $z'$ magnitudes, 
whose rest-frame wavelengths are 
$1580$ {\AA} and $2190$ {\AA} for $z=3.1$
and 
$1390$ {\AA} and $1930$ {\AA} for $z=3.7$.
For the LBGs at $z \sim 4$, $6$, and $7$, we take the values given in
\cite{bouwens2009b,bouwens2010b}. 
In addition, we calculate $\beta=-2.9 \pm 0.4$ for the $z \sim 7$ stacked LBG 
from its $J$ and $H$ magnitudes\footnote{The rest-frame wavelengths are $1580$ {\AA} and $1950$ {\AA}.}
given in \cite{labbe2010}.

We see a weak correlation in the LBGs of individual redshifts
that fainter objects have smaller $\beta$, i.e., bluer spectra,
and a tendency that 
$\beta$ at fixed $M_{\rm UV}$ appears to become smaller with redshift.
A similar correlation can be seen for the LAEs at $z=3.1$ and $3.7$
if we assume that evolution between these redshifts is negligible
and treat them collectively as $z \sim 3.5$ objects.
In comparison with this possible correlation seen in LAEs at
$z \sim 3.5$, the LAEs at $z=5.7$ and $6.6$ are offset
toward smaller $\beta$,
although the significance is at most $1 \sigma$ levels due to the
large uncertainties in the $\beta$ measurement especially at $z=6.6$.
This offset, if real, 
might
suggest that the stellar populations of
typical LAEs at $z \sim 6$ -- $7$ are younger, more metal-poor, and/or
with less dust than those at $z \sim 3.5$.

Moreover, the LAEs at $z = 5.7$ and $6.6$ are slightly below
the $\beta$ - $M_{\rm UV}$ correlation of $z \sim 6$ and $7$ LBGs.
Although the difference between LAEs and LBGs at $z\sim 6-7$
is within $1\sigma$ uncertainties, it 
might
imply that
the stellar population is slightly different between these galaxies.
\cite{shapley2003} have reported that the UV spectra of $z \sim 3$
LBGs become bluer with increasing Ly$\alpha$ equivalent width
from large negative values (strong absorption) to large positive
values \citep[see also,][]{kornei2010}.
Similar trends have also been found for $z\sim 4$ LBGs
by \cite{pentericci2007}, \cite{vanzella2009},
and \cite{stark2010}.
This trend 
might
continue to $z\sim 6 -7$,
since there 
seems to be
a hint of systematically bluer continua
for LAEs than LBGs in Figure \ref{fig:betas}.

\section{SED FITTING} \label{sec:sedfitting}

Since LAEs tend to be young star-forming galaxies
\citep[e.g.,][]{gawiser2007,pirzkal2007,lai2008,ono2009},
it is worthwhile to consider nebular emission
in population synthesis modeling.
Here we make model SEDs in two extreme cases:
$f_{\rm esc}^{\rm ion} = 0$ where ionizing photons are totally converted
into nebular emission,
and $f_{\rm esc}^{\rm ion} = 1$ where all ionizing photons escape from
the galaxy.
We call the former the $''$stellar $+$ nebular$''$ case,
and the latter the $''$pure stellar$''$ case.
We calculate nebular spectra (lines and continua)
basically following the manner given in \cite{schaerer2009}.

We use the stellar population synthesis model of GALAXEV
\citep[][hereafter BC03]{bc2003} to produce stellar SEDs\footnote{We do not use new population synthesis models 
which include thermally pulsating aymptotic giant branch (TP-AGB) 
stars \citep[e.g.,][]{maraston2005,bruzual2007}, 
because LAEs tend to be very young and 
the contribution of TP-AGB stars should be negligible \citep[e.g.,][]{ono2009}. },
adopting Salpeter's initial mass function \citep{salpeter1955}
with lower and upper mass cutoffs of $0.1$ and $100 M_\odot$.
We assume constant star formation history\footnote{Since 
most LAEs are very young, 
constant star formation history (SFH) is a reasonable approximation. 
Younger ages will be obtained if exponentially decaying SFH is 
assumed, while older ages will be obtained for 
smoothly-rising SFH, which has been recently applied 
by several authors \citep[e.g.,][]{stark2009}.} 
and consider two stellar metallicities $Z = 0.2 Z_\odot$ and $0.02 Z_\odot$.

Nebular emission is calculated
under the assumption of electron temperature $T_{\rm e} = 10^4$ K,
electron density $n_{\rm e} = 10^2$ cm$^{-3}$,
and case B recombination.
We include H recombination lines from Balmer, Paschen, and Brackett
series\footnote{We do not include Ly$\alpha$ line, 
since Ly$\alpha$ photons are resonantly scattered
by neutral hydrogen and its strength is quite uncertain. 
Instead, 
we derive the Ly$\alpha$-free continuum magnitude $m_{\rm con}$ 
from $i'$- and $NB816$-band photometry for the $z=5.7$ LAE, 
and $z'$- and $NB921$-band photometry for the $z=6.6$ LAE, 
as described in Section \ref{subsec:photometry}.}.
We calculate H$\beta$ line luminosity by \citep[e.g.,][]{osterbrock2006}
\begin{equation}
L ({\rm H}\beta) \, [{\rm erg} \, {\rm s}^{-1}]
       = 4.78 \times 10^{-13} \left( 1 - f_{\rm esc}^{\rm ion} \right) N_{\rm Lyc},
\end{equation}
where $N_{\rm Lyc}$ is the number of ionizing photons produced
per second.
We do not consider absorption of ionizing photons by dust.
The luminosities of the other H recombination lines are computed
from $L ({\rm H}\beta)$ using the table of relative intensities
of these lines given in \cite{storey1995}.
We also include nebular lines from non-hydrogens using the empirical
relative line intensities compiled by \cite{anders2003},
on the assumption that the gaseous metallicity is equal to
the stellar metallicity.

Nebular continuum emission is calculated by
\citep[e.g.,][]{krueger1995}
\begin{equation}
L_\nu
       = \frac{\gamma_\nu^{\rm (total)}}{\alpha_{\rm B}}
               \left( 1 - f_{\rm esc}^{\rm ion} \right) N_{\rm Lyc},
\end{equation}
where $\alpha_{\rm B}$ is the case B recombination coefficient
for hydrogen.
The continuum emission coefficient $\gamma_\nu^{\rm (total)}$,
considering free-free and free-bound emission by H, neutral He,
and singly ionised He, as well as the two-photon continuum of H,
is given by
\begin{equation}
\gamma_\nu^{\rm (total)}
       = \gamma_\nu^{\rm (HI)} + \gamma_\nu^{\rm (2q)}
               + \gamma_\nu^{\rm (HeI)} \frac{n ({\rm He}^+)}{n ({\rm H}^+)}
               + \gamma_\nu^{\rm (HeII)} \frac{n ({\rm He}^{++})}{n ({\rm H}^+)}.
\end{equation}
The emission coefficients $\gamma_\nu^{\rm (i)}$ ($i=$ HI, 2q, HeI, HeII)
for wavelengths below and above $1\mu$m are taken from
tables 4 -- 9 of \cite{aller1984} and
tables I and II of \cite{ferland1980},
respectively\footnote{We assume $\gamma_\nu^{\rm (2q)} = 0$ and
$\gamma_\nu^{\rm (HeI)} = \gamma_\nu^{\rm (HI)}$
at $\lambda \geq 1\mu$m \citep[e.g.,][]{schaerer1998}.}.
The abundance ratios are set to be
$n ({\rm He}^+) / n ({\rm H}^+) = 0.1$ and
$n ({\rm He}^{++}) / n ({\rm H}^+) = 0$
\citep[e.g.,][]{brown1970,krueger1995}.

For the dust extinction of output stellar spectra, we use Calzetti's
extinction law \citep{calzetti2000} and vary $E(B-V)_\star$
as a free parameter over $0$ and $1.50$ with an interval of $0.01$.
For the dust extinction of nebular emission,
we assume $E(B-V)_{\rm gas} = E(B-V)_\star$ as proposed
by \cite{erb2006c}\footnote{
We find that adopting $E(B-V)_{\rm gas} = E(B-V)_\star/0.44$,
which is proposed by \cite{calzetti2000}, does not significantly
change our results.}.

Note that we do not consider the effect of dust extinction on
$f_{\rm esc}^{\rm ion}$.
In other words, we assume that a Lyman continuum photon either
ionizes a neutral hydrogen atom or escapes into the intergalactic
medium through, e.g., holes in the gas.
Dust extinction for Lyman continuum emission can substantially
reduce the hydrogen-ionizing flux \citep[e.g.,][]{inoue2001c}.
However, the exact level of Lyman continuum extinction
is difficult to assess even for galaxies in the local Universe,
and much more so at higher redshifts \citep[e.g.,][]{zackrisson2008},
although some authors suggest that the effect is very small
for high-redshift galaxies \citep[e.g.,][]{gnedin2008,razoumov2010}

In Section \ref{subsec:Implication_for_fesc}, we place upper limits to $f_{\rm esc}^{\rm ion}$
from the minimum luminosity of nebular emission required to
reproduce the observed SEDs.
Those are regarded as conservative upper limits,
since a non-zero fraction of Lyman continuum photons not converted
into nebular emission will in practice be absorbed by dust
before escaping into the IGM.

We perform the standard SED fitting method
\citep[for details, see Section 3 of ][]{ono2009}.
We make a large set of stellar-mass-normalized model SEDs,
varying age and dust extinction.
We then redshift them to $z=5.70$ and $6.56$
and convolve them with bandpasses to calculate flux densities.
For each object, we search for the best-fitting SED that
minimizes $\chi^2$ separately for $f_{\rm esc}^{\rm ion}=0$ and 1
and separately for $Z=0.2 Z_\odot$ and $0.02 Z_\odot$.
Since 
stellar mass 
$M_{\rm star}$ is the amplitude of a model SED,
we obtain the best-fitting $M_{\rm star}$ by
solving $\partial \chi^2 / \partial M_{\rm star} = 0$.
The errors in the best-fit SED parameters
correspond to the $1\sigma$ confidence interval
($\Delta\chi^2 < 1$) for each parameter.

Since our $i'$-band flux density for the $z=5.7$ LAE
and $z'$-band flux density for the $z=6.6$ LAE
are contaminated from Ly$\alpha$ emission and 
IGM absorption,
we use $m_{\rm con}$ calculated in Section \ref{subsec:photometry}
instead of these flux densities.
We do not use short wavebands (i.e., $BVR$ for $z=5.7$
and $BVRi'$ for $z=6.6$), since they suffer from the IGM absorption
shortward of the Ly$\alpha$ wavelength, the amount of which
considerably differs among the lines of sight.
Thus, model SEDs are fitted to the observed flux densities
in $m_{\rm con}$, $z'$, $J$, $H$, $K$, $3.6\mu$m, $4.5\mu$m,
$5.8\mu$m, and $8.0\mu$m-band for $z=5.7$,
and to the same bands except $z'$ for $z=6.6$.
The free parameters in the fitting are stellar mass, age, and
dust extinction.
The degrees of freedom are six for $z=5.7$ and five for $z=6.6$.

\begin{figure*}
 \begin{center}
  \includegraphics[scale=1.0]{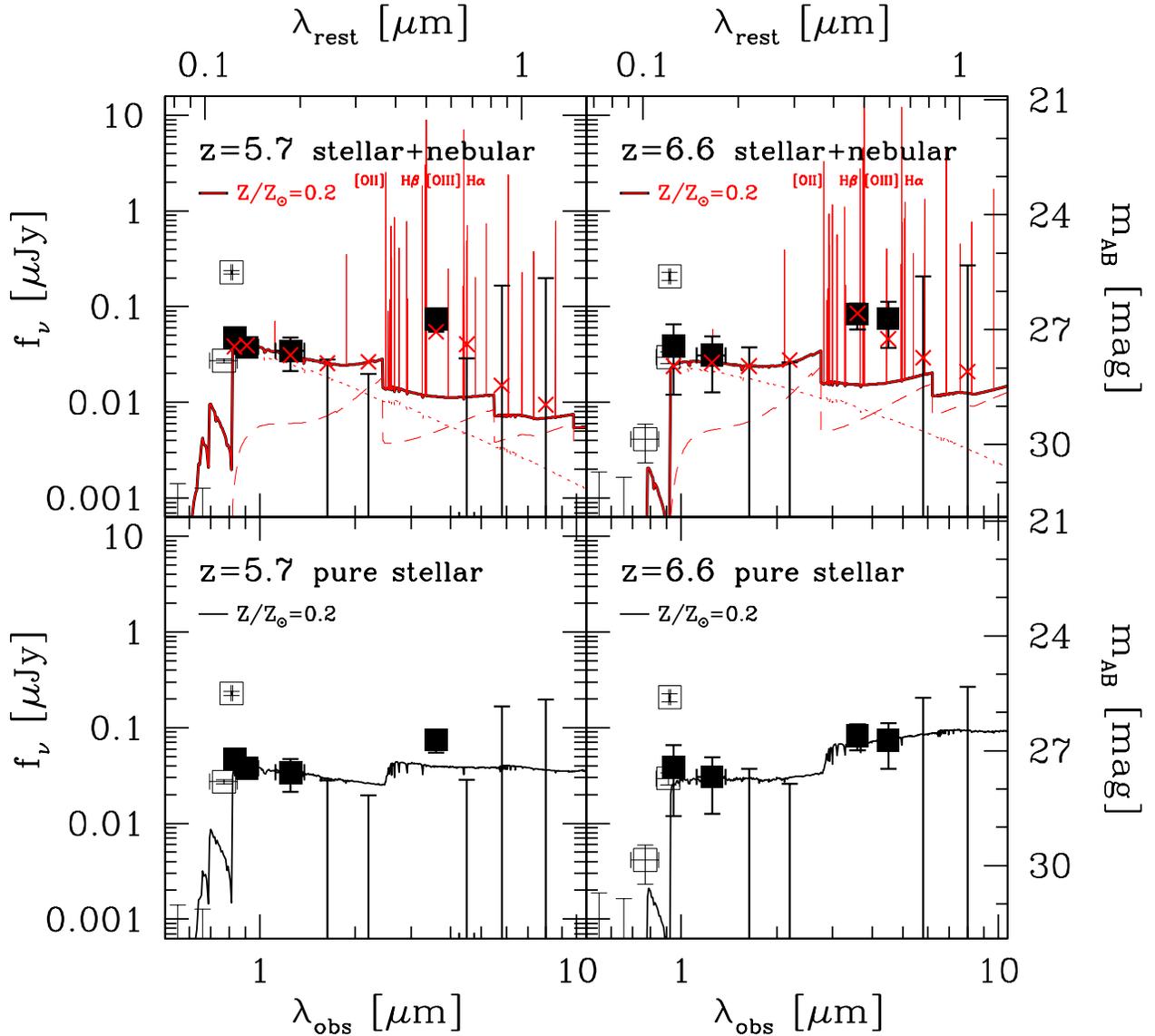}
\end{center}
 \caption[]
{
Best-fit SEDs (curves) and observed magnitudes (filled squares:
used for SED fitting, open squares: not used) for the $z=5.7$
LAE (left panels) and the $z=6.6$ LAE (right panels).
The top panels are for {\lq}stellar $+$ nebular{\rq}
($f_{\rm esc}^{\rm ion}=0$) models.
The red solid curves indicate the best-fit SEDs that are the
sum of a stellar SED (red dotted curve) and a nebular
SED (red dashed curve).
The crosses indicate synthesized flux densities in individual
bandpasses.
The bottom panels are for {\lq}pure stellar{\rq}
($f_{\rm esc}^{\rm ion}=1$) models.
}
\label{fig:SEDs}
\end{figure*}

\begin{deluxetable*}{cccccccc} 
\tablecolumns{8} 
\tablewidth{0pt} 
\tablecaption{SED Fitting Results for the LAEs at $z=5.7$ and $6.6$ \label{tab:SEDfitting}}
\tablehead{
    \colhead{model}    & \colhead{$Z$} & \colhead{$\log M_{\rm star}$}    & \colhead{$E(B-V)_\star$} & \colhead{$\log$(Age)}    
    & \colhead{$\log$(SFR)}    & \colhead{$\log$(SSFR)}  & \colhead{$\chi^2$} \\
    \colhead{}    & \colhead{[$Z_\odot$]}    & \colhead{$[M_\odot]$}    & \colhead{[mag]} & \colhead{[yr]}    
    & \colhead{[$M_\odot$ yr$^{-1}$]}  & \colhead{[yr$^{-1}$]}  & \colhead{}
}
\startdata 
\multicolumn{8}{c}{$z=5.7$ LAE} \\ 
\hline \\
stellar $+$ nebular
& $0.2$  & $7.49^{+0.20}_{-0.03}$ &  $0.00^{+0.03}_{-0.00}$  
&  $6.50^{+0.62}_{-1.40}$  &  $0.99^{+1.48}_{-0.46}$  &  $-6.50^{+1.40}_{-1.78}$  &  $7.75$ 
\\
pure stellar
& $0.2$  & $8.69^{+0.32}_{-0.56}$ &  $0.00^{+0.07}_{-0.00}$  
&  $8.46^{+0.35}_{-0.91}$  &  $0.31^{+0.42}_{-0.02}$  &  $-8.38^{+1.80}_{-0.48}$  &  $10.3$ 
\\
stellar $+$ nebular
& $0.02$ & $7.49^{+0.37}_{-0.01}$ &  $0.00^{+0.05}_{-0.00}$  
&  $6.62^{+0.58}_{-1.52}$  &  $0.87^{+1.63}_{-0.33}$  &  $-6.62^{+1.52}_{-2.13}$  &  $9.00$ 
\\ 
pure stellar
& $0.02$ & $8.88^{+0.26}_{-0.39}$ &  $0.00^{+0.07}_{-0.00}$  
&  $8.71^{+0.25}_{-0.55}$  &  $0.26^{+0.35}_{-0.01}$  &  $-8.61^{+1.13}_{-0.24}$  &  $10.4$ 
\\
\hline \\
 \multicolumn{8}{c}{$z=6.6$ LAE} \\ 
 \hline \\ 
stellar $+$ nebular
& $0.2$ & $7.95^{+1.66}_{-0.30}$ &  $0.11^{+0.24}_{-0.11}$  
&  $5.95^{+2.96}_{-0.85}$  &  $2.00^{+1.21}_{-1.66}$  &  $-5.95^{+0.85}_{-2.86}$  &  $2.53$ 
\\
pure stellar
& $0.2$ & $9.43^{+0.28}_{-0.33}$ &  $0.09^{+0.19}_{-0.09}$  
&  $8.91^{+0.00}_{-0.95}$  &  $0.63^{+0.71}_{-0.24}$  &  $-8.81^{+2.33}_{-0.00}$  &  $2.86$ 
\\ 
stellar $+$ nebular
& $0.02$ & $8.07^{+1.63}_{-0.30}$ &  $0.13^{+0.25}_{-0.11}$  
&  $6.38^{+2.53}_{-1.28}$  &  $1.69^{+1.63}_{-1.27}$  &  $-6.38^{+1.28}_{-2.42}$  &  $2.46$ 
\\ 
pure stellar
& $0.02$ & $9.57^{+0.28}_{-0.32}$ &  $0.14^{+0.20}_{-0.11}$  
&  $8.91^{+0.00}_{-0.85}$  &  $0.77^{+0.74}_{-0.28}$  &  $-8.80^{+2.26}_{-0.00}$  &  $3.05$ 
\enddata 
\end{deluxetable*} 

\begin{figure}
 \begin{center}
  \includegraphics[scale=0.8]{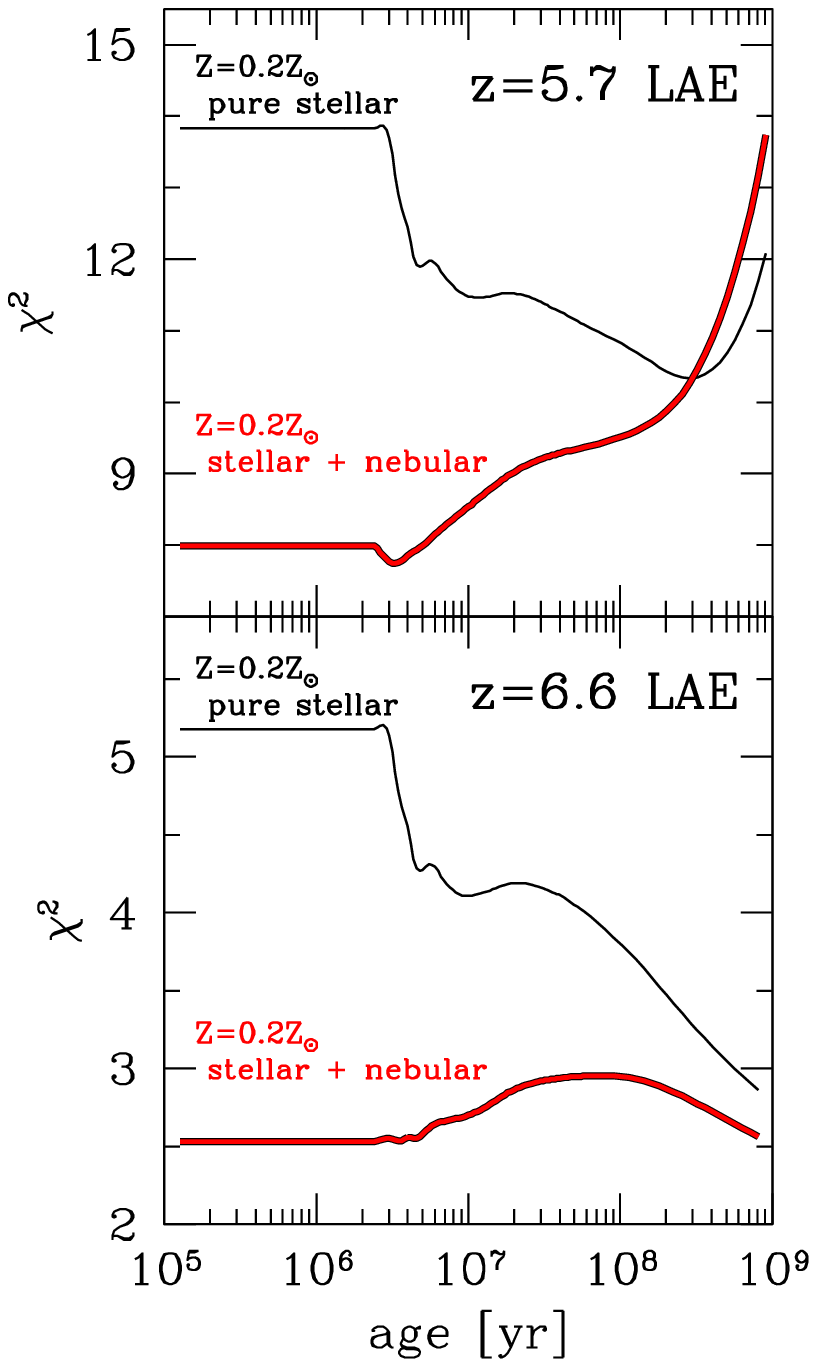}
\end{center}
 \caption[]
{
$\chi^2$ of the best-fit model as a function of age
(top: $z=5.7$ LAE, bottom: $z=6.6$ LAE).
The red curves are for $f_{\rm esc}^{\rm ion} = 0$ and the black curves
for $f_{\rm esc}^{\rm ion} = 1$.
}
\label{fig:age_chi2}
\end{figure}

\section{RESULTS AND DISCUSSION} \label{sec:results_and_discussion}

\subsection{Stellar Populations of Lyman Alpha Emitters \\ at $z \sim 6$ -- $7$}\label{subsec:stellarpop_of_LAEs}

Table \ref{tab:SEDfitting} summarizes the results of the SED
fitting for our LAEs.
First, we find that in both $f_{\rm esc}^{\rm ion}=0$ and $f_{\rm esc}^{\rm ion}=1$
the best-fit models for $Z=0.2 Z_\odot$ and $0.02 Z_\odot$ are
very similar to each other.
This implies that the changes in the stellar and nebular emission
spectra over $0.02 \lesssim Z/Z_\odot \lesssim 0.2$ are not
large enough to significantly alter the best-fit parameters.
In what follows we concentrate on the results for $Z=0.2 Z_\odot$
for simplicity.

Next, we find that the best-fit models are extremely different
between $f_{\rm esc}^{\rm ion}=0$ and 1.
For $f_{\rm esc}^{\rm ion}=0$ the best-fit models have relatively small
stellar masses, $3 \times 10^7 M_\odot$ ($z=5.7$) and
$1 \times 10^8 M_\odot$ ($z=6.6$), and young ages,
3 Myr ($z=5.7$) and 1 Myr ($z=6.6$).
On the other hand, for $f_{\rm esc}^{\rm ion}=1$ the stellar masses are
more than one order of magnitude higher, $5 \times 10^8 M_\odot$
and $3 \times 10^9 M_\odot$, and the ages are more than two
orders of magnitude older, 300 Myr and 800 Myr.
As we see below, these great differences are related to how to
account for the observed bright IRAC magnitudes, or equivalently
the red UV-to-optical colors.
In contrast to the stellar mass and age, the dust extinction is
consistent with $E(B-V)_\star = 0$ in both cases for both objects,
suggesting that typical LAEs at $z \sim 6$ -- 7 are nearly
free from dust extinction.

Figure \ref{fig:SEDs} compares the best-fit model spectra with
the observed flux densities.
The bottom panels show the results for $f_{\rm esc}^{\rm ion} =1$,
the $''$pure stellar$''$ case.
For $z=5.7$, the best-fit model matches the observation
at $\lambda_{\rm obs} \lesssim 2 \mu$m but undershoots
the data point at $3.6\mu$m.
This offset will be reduced if older ages or larger $E(B-V)_\star$
values are adopted, but such models will then not fit the blue
UV spectra.
The observed SED of the $z=6.6$ LAE resembles that of the
$z=5.7$ LAE, with blue UV colors and detection at $3.6\mu$m.
In addition, the $z=6.6$ LAE has a relatively bright flux density
at $4.5\mu$m as well although less than $3\sigma$ detection.
The overall shape of the $z=6.6$ SED is reproduced well by
a pure stellar model.

The top panels of Figure \ref{fig:SEDs} show the results
for $f_{\rm esc}^{\rm ion}=0$, i.e., the $''$stellar $+$ nebular$''$ case.
We find that for $z=5.7$ the discrepancy at $3.6\mu$m seen
in the bottom panel almost disappears thanks to strong nebular
emission lines such as {\sc [Oiii]} and H$\beta$ contributing
to this bandpass; the stellar emission has only a minor
contribution.
Thus, adopting a very young stellar population with strong
nebular emission can simultaneously fit the observed
blue UV color and the red UV-to-optical color.
For $z=6.6$ as well, the best-fit model with $f_{\rm esc}^{\rm ion} = 0$
reproduces the observed SED well in a similar manner.

Then, which of $f_{\rm esc}^{\rm ion}=1$ and $f_{\rm esc}^{\rm ion} =0$ is more
favored?
For $z=5.7$ it is easy to answer this question.
As shown in Table \ref{tab:SEDfitting}, the best-fit $\chi^2$
for $f_{\rm esc}^{\rm ion}=0$, $7.75$, is significantly smaller than that
for $f_{\rm esc}^{\rm ion}=1$, $10.3$.
Indeed, the $f_{\rm esc}^{\rm ion}=0$ model fits the $3.6\mu$m data point
and blue $3.6\mu$m -- $4.5\mu$m color far better than
the $f_{\rm esc}^{\rm ion}=1$ model.
In addition, the small photometric errors in the $m_{\rm con}$, $z'$,
and $J$ magnitudes do not permit old-age, high-mass models which
fit the $3.6\mu$m data but instead give red UV colors.

For $z=6.6$, on the other hand, the best-fit $f_{\rm esc}^{\rm ion}=1$ model
gives almost the same $\chi^2$ ($2.9$) as the $f_{\rm esc}^{\rm ion}=0$ model
($2.5$), and we cannot conclude which is favored solely from
the $\chi^2$ values.
This is partly because the larger photometric errors in the $J$
and shorter bandpasses, and flatter $3.6\mu$m -- $4.5\mu$m color
permit old, massive models with red UV colors.
However, typical LAEs at $z \sim 3$ -- $5$ have been consistently
found to be young and low-mass objects
\citep[e.g.,][]{gawiser2007,pirzkal2007,lai2008,ono2009},
and our study shows that this trend continues at $z=5.7$.
Combining this fact with an argument that it seems unnatural
that LAEs at earlier epochs are {\it older} and {\it more massive},
we take the model with $f_{\rm esc}^{\rm ion}=0$ for the $z=6.6$ LAE\footnote{We 
cannot completely rule out the possibility that
due to some selection effect unique to the $z=6.6$ sample,
our $z=6.6$ LAEs are biased toward very massive objects.
However, such selection effects are very unlikely,
since the $z=6.6$ sample has been selected in a similar manner
to the $z=5.7$ sample from the same data set,
and the stacked objects at these two redshifts have
similar $L({\rm Ly}\alpha)$ and $M_{\rm UV}$,
as found in Table \ref{tab:photo_prop}.}.

We conclude that typical LAEs at $z \sim 6$ -- $7$ have 
low stellar masses of $\sim \left( 3-10 \right) \times 10^{7}M_\odot$,
very young ages of $\sim 1-3$ Myr,
and negligible dust extinction.
We thus propose that they are candidates of galaxy building blocks 
at an early stage of galaxy formation.
This proposal can be regarded as an extension toward higher 
redshift of a similar idea which has been proposed for low-redshift 
LAEs based on apparent magnitudes and sizes 
(\citealt{pascarelle1996} for $z \sim 2$; 
see also \citealt{ouchi2003} who found that $z\sim5$ LAEs are UV faint)
and stellar population analysis (\citealt{gawiser2007} for $z \sim 3$).
This proposal is also consistent with low dark-halo masses 
estimated for LAEs from clustering analysis  
(\citealt{gawiser2007} for $z\sim 3$, 
\citealt{guaita2009} for $z\sim 2$, 
\citealt{ouchi2010} for $z \sim 3$ -- $7$).

\subsection{Constraints on Ly$\alpha$ Escape Fraction}

We estimate for our LAEs the Ly$\alpha$ escape fraction,
$f_{\rm esc}^{{\rm Ly}\alpha}$, by:
\begin{equation}
f_{\rm esc}^{{\rm Ly}\alpha}
 = \frac{L_{\rm obs}({\rm Ly}\alpha) }{ L_{\rm int}({\rm Ly}\alpha)},
\end{equation}
where $L_{\rm obs}({\rm Ly}\alpha)$ is the observed Ly$\alpha$
luminosity and $L_{\rm int}({\rm Ly}\alpha)$ is the intrinsic
Ly$\alpha$ luminosity computed from the SFR on the assumption of
case B using
$L_{\rm int}({\rm Ly}\alpha)$ [erg s$^{-1}$]
$= 1.1 \times 10^{42}$ SFR [$M_\odot$ yr$^{-1}$]
\citep{brocklehurst1971,kennicutt1998}.
As is easily noticed, $f_{\rm esc}^{{\rm Ly}\alpha}$ corresponds
to the fraction of Ly$\alpha$ photons produced in the galaxy
which escape from absorption by the galaxy's ISM and absorption
by the IGM at the galaxy's redshift.
Because the dust extinction of our LAEs is negligibly small
(Section \ref{subsec:stellarpop_of_LAEs}),
Ly$\alpha$ photons propagating in the ISM are mostly
just scattered by HI gas without absorption.
In this case, $f_{\rm esc}^{{\rm Ly}\alpha}$ depends mostly
on the strength of the IGM absorption.

\begin{figure}
 \begin{center}
  \includegraphics[scale=0.8]{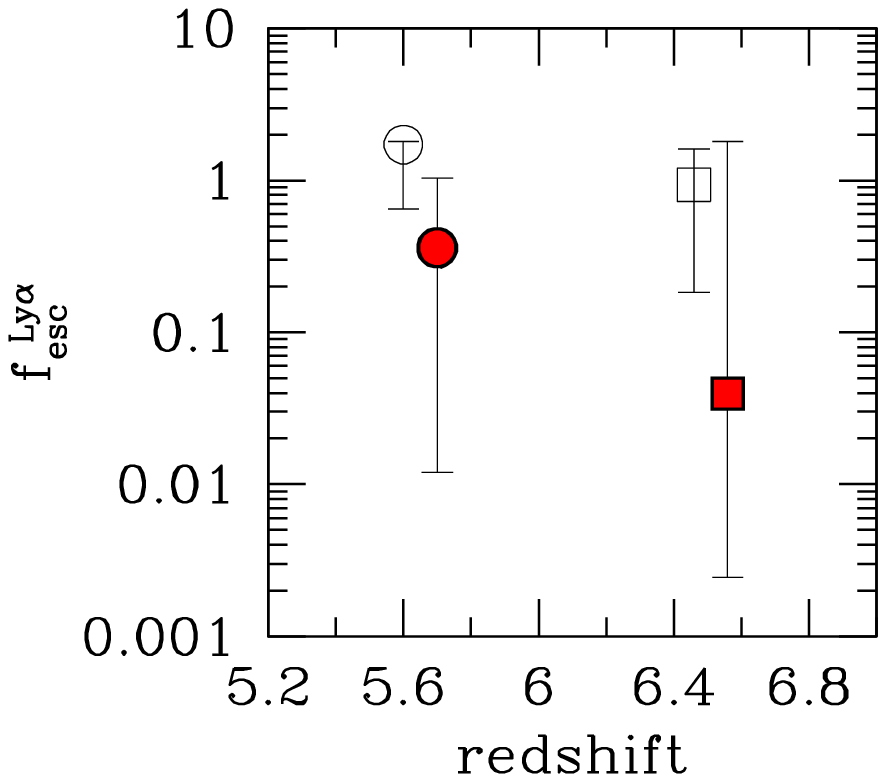}
\end{center}
 \caption[]
{
Ly$\alpha$ escape fraction as a function of redshift.
The circles (squares) correspond to $z=5.7$ ($6.6$).
The filled and open symbols are for $f_{\rm esc}^{\rm ion} = 0$ and $1$,
respectively, where the open symbols have been shifted by $-0.1$
along the $x$-axis for clarity. 
Although $f_{\rm esc}^{\rm ion} = 1$ is physically 
unrealistic because Ly$\alpha$ photons cannot be produced, 
we plot the results to show how $f_{\rm esc}^{\rm Ly\alpha}$ 
varies with $f_{\rm esc}^{\rm ion}$ in our method.}
\label{fig:fesc_redshift}
\end{figure}

Substituting the observed Ly$\alpha$ luminosity and SFR derived
from our SED fitting,
we obtain $f_{\rm esc}^{{\rm Ly}\alpha} \simeq 0.36^{+0.68}_{-0.35}$ for $z=5.7$ LAEs,
and $f_{\rm esc}^{{\rm Ly}\alpha} \simeq 0.040^{+1.8}_{-0.038}$ for $z=6.6$ LAEs,
as plotted in Figure \ref{fig:fesc_redshift}.
We find a tentative decrease in $f_{\rm esc}^{{\rm Ly}\alpha}$
from $z=5.7$ to $6.6$. 
\cite{ono2009} have found that 
LAEs at $z \sim 3$ -- $4$ have 
$f_{\rm esc}^{{\rm Ly}\alpha} \simeq 0.1$ -- $1$.
\cite{hayes2010b} have obtained the median value 
of $f_{\rm esc}^{{\rm Ly}\alpha}$ for $z=2.2$ LAEs 
to be higher than $0.32$.
These results might suggest that 
the Ly$\alpha$ escape fraction of LAEs is nearly constant 
up to $5.7$, and then decreases toward $z=6.6$. 
This decrease, if real, could be due to an increase with redshift
in the hydrogen neutral fraction of the IGM, thereby
Ly$\alpha$ photons are scattered more frequently.
Indeed, a possible increase in the neutral fraction has been
proposed from an observed decline in the number density of bright
LAEs from $z=5.7$ to $z=6.6$ by \cite{kashikawa2006}.

Further discussion on the neutral fraction is, however, difficult
given the quality of the $f_{\rm esc}^{{\rm Ly}\alpha}$ measurements.
Note also that there may be some other mechanisms which change
$f_{\rm esc}^{{\rm Ly}\alpha}$.
For example, unlike our assumption, Ly$\alpha$ emissivity at
a given SFR might decrease from $z=5.7$ to $6.6$.
Although Figure \ref{fig:fesc_redshift} shows potential
usefulness of $f_{\rm esc}^{{\rm Ly}\alpha}$) as a means to evaluate
the IGM neutral fraction near the reionization epoch,
much deeper multiband photometry and more detailed SED modeling
will be needed to obtain a reliable constraint.

\begin{figure}
 \begin{center}
  \includegraphics[scale=0.85]{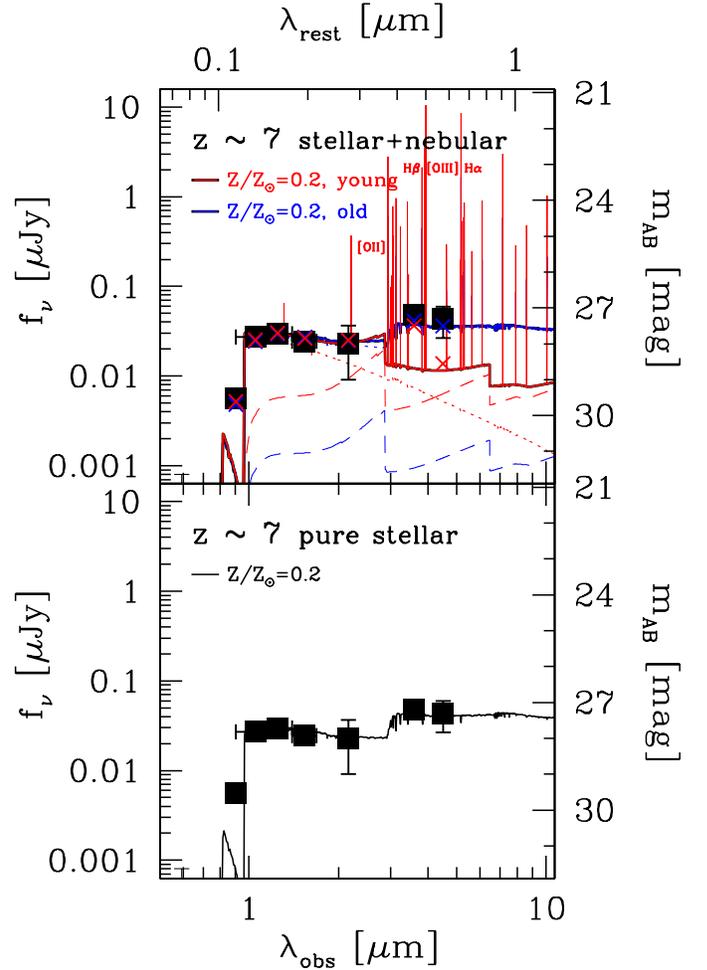}
\end{center}
 \caption[]
{
Same as Figure \ref{fig:SEDs}, but for
the $z \sim 7$ $z$-dropout galaxy \citep{labbe2010}.
In the top panel,
two equally well fitted $f_{\rm esc}^{\rm ion}=0$ models are plotted:
the {\lq}young{\rq} model in red and the {\lq}old{\rq} model in blue.
}
\label{fig:SEDs_dropout}
\end{figure}

\begin{deluxetable*}{cccccccc} 
\tablecolumns{8} 
\tablewidth{0pt} 
\tablecaption{SED Fitting Results for the $z \sim 7$ $z$-dropout galaxy \label{tab:SEDfitting_zdrop}}
\tablehead{
    \colhead{model}    & \colhead{$Z$} & \colhead{$\log M_{\rm star}$}    & \colhead{$E(B-V)_\star$} & \colhead{$\log$(Age)}    
    & \colhead{$\log$(SFR)}    & \colhead{$\log$(SSFR)}  & \colhead{$\chi^2$} \\
    \colhead{}    & \colhead{[$Z_\odot$]}    & \colhead{$[M_\odot]$}    & \colhead{[mag]} & \colhead{[yr]}    
    & \colhead{[$M_\odot$ yr$^{-1}$]}  & \colhead{[yr$^{-1}$]}  & \colhead{}
}
\startdata 
 \multicolumn{8}{c}{$z \sim 7$ $z$-dropout} \\ 
 \hline \\ 
\multirow{2}{*}{stellar $+$ nebular}
& \multirow{2}{*}{$0.2$}
& $7.72^{+0.10}_{-0.18}$ &  $0.04^{+0.03}_{-0.04}$  &  $6.16^{+0.52}_{-1.06}$  &  $1.56^{+1.16}_{-0.68}$  &  $-6.16^{+1.06}_{-0.94}$  &  $7.35$ 
\\ 
 
&
& $8.83^{+0.22}_{-0.30}$ &  $0.00^{+0.01}_{-0.00}$  
&  $8.61^{+0.25}_{-0.35}$  &  $0.31^{+0.05}_{-0.01}$  &  $-8.52^{+0.67}_{-0.24}$  &  $6.90$ 
\\
pure stellar
& $0.2$  & $8.98^{+0.12}_{-0.27}$ &  $0.00^{+0.01}_{-0.00}$  
&  $8.76^{+0.10}_{-0.30}$  &  $0.32^{+0.05}_{-0.00}$  &  $-8.66^{+0.43}_{-0.10}$  &  $6.26$ 
\enddata 
\tablecomments{
For the $z \sim 7$ $z$-dropout, shown are two equally well fitted
$f_{\rm esc}^{\rm ion} = 0$ models, a very young model and a very old model.
See text for details.
}
\end{deluxetable*} 

\begin{figure}
 \begin{center}
  \includegraphics[scale=0.8]{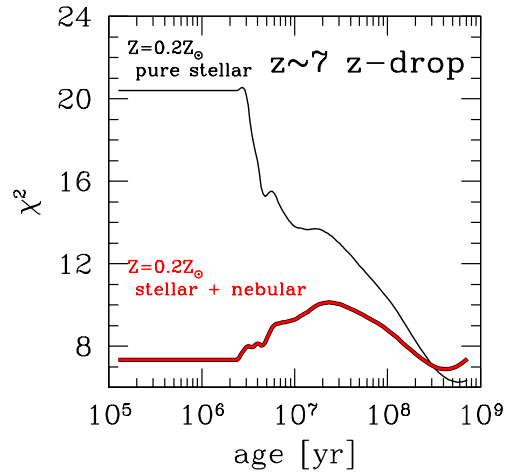}
\end{center}
 \caption[]
{
Same as Figure \ref{fig:age_chi2}, but for the $z \sim 7$ $z$-dropout
galaxy \citep{labbe2010}.
}
\label{fig:age_chi2_zdrop}
\end{figure}

\begin{figure}
 \begin{center}
  \includegraphics[scale=0.8]{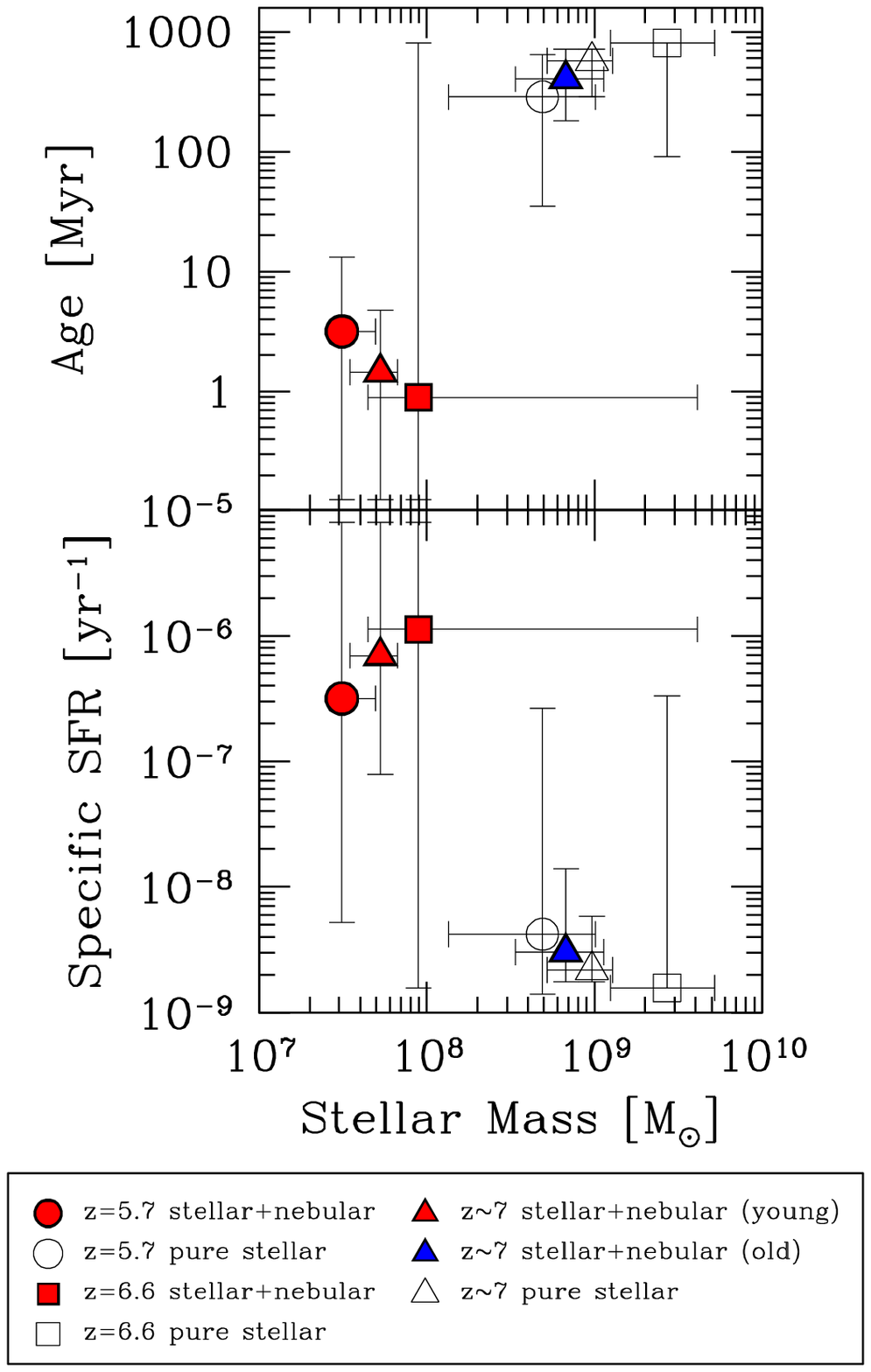}
\end{center}
 \caption[]
{
\textbf{Top panel}:
age versus stellar mass $M_{\rm star}$.
The circles, squares, and triangles are for the $z=5.7$ LAE,
$z=6.6$ LAE, and $z\sim 7$ dropout galaxy \citep{labbe2010},
respectively.
The results for $f_{\rm esc}^{\rm ion} = 0$ are colored in red, while
those for $f_{\rm esc}^{\rm ion} = 1$ are shown in open symbols.
For the $z\sim 7$ dropout galaxy,
we also plot the old $f_{\rm esc}^{\rm ion} = 0$ model in blue.
\textbf{Bottom panel}:
specific star formation rate (SSFR $=$ SFR$/M_{\rm star}$) versus
stellar mass $M_{\rm star}$.
The symbols are the same as in the top panel.
}
\label{fig:Mstar_sSFR_age}
\end{figure}

\subsection{Comparison with LBGs at $z \sim 7$}\label{subsec:stellarpop_of_LBGs}

We fit model SEDs to a $z\sim7$ $z$-dropout composed of $14$ objects
recently discovered by the HST/WFC3 survey \citep{oesch2010,labbe2010},
to compare with our LAEs. 
They have been selected 
from an extremely deep ($5\sigma \approx 28.6-28.7$ 
over $0.''25$-radius apertures 
in $Y_{105}$, $J_{125}$, and $H_{160}$) 
area of $4.7$ arcmin$^2$  
using $z_{850} - Y_{105}$ and $Y_{105} - J_{125}$ colors 
and have an expected redshift distribution $z \sim 6.5 - 7.5$. 
We consider the observed flux densities
in the $z_{850}$, $Y_{105}$, $J_{125}$, $H_{160}$, $K$, $3.6\mu$m,
$4.5\mu$m band taken from Table 1 of \cite{labbe2010},
assuming a redshift of $z = 6.88$, which is derived by
\cite{labbe2010}.
This stacked object is similarly faint in the rest UV continuum 
($J_{125} = 29.6$) to our LAEs, and thus suitable for comparison.
We do not use two $z$-dropout candidates reported by 
\cite{capak2009} and 
$11$ $z_{850 {\rm LP}}$-dropouts discovered by \cite{gonzalez2009}, 
because they are too bright 
(the former have $J \sim 23$ 
and the latter $J_{110{\rm W}} \sim 26 - 27.5$).
In our SED fitting, 
we rule out the $i_{775}$-band data
although \cite{labbe2010} used it for their SED fitting,
since it suffers from the IGM absorption shortward of Ly$\alpha$
wavelength,
and the amount of the absorption differs with the line of sight.
The free parameters are thus stellar mass, age, and dust extinction,
and the degrees of freedom are four.

This object has also been SED-fitted by \cite{schaerer2010},
using stellar population synthesis models with nebular emission.
The major difference between their and our modelings are the star
formation history assumed.
\cite{schaerer2010} have assumed an exponentially declining star
formation rate with its $e$-folding time as a free parameter,
while we assume a constant star formation rate.
For a fair comparison with our results on LAEs, we reanalyze
the $z$-dropout SED with our models.
The best-fit parameters are given in Table \ref{tab:SEDfitting}.

The bottom panel of Figure \ref{fig:SEDs_dropout}
shows the result for $f_{\rm esc}^{\rm ion}=1$.
We find that the observed SED is well explained by
a low-mass ($\log \left( M_{\rm star} [M_\odot] \right) = 8.98^{+0.12}_{-0.27}$)
and moderately aged ($\log \left({\rm Age [yr]} \right) = 8.76^{+0.10}_{-0.30}$) model
with little dust extinction ($E(B-V)_\star = 0.00^{+0.01}_{-0.00}$)\footnote{The
age suggests that the formation redshift of this object is $\geq 9.5$.
}.
This result is consistent with that of \cite{labbe2010}
who examined $f_{\rm esc}^{\rm ion}=1$ models alone.

The top panel of Figure \ref{fig:SEDs_dropout}
shows the result for $f_{\rm esc}^{\rm ion}=0$.
We obtain almost the same parameters as the $f_{\rm esc}^{\rm ion}=1$ model:
$\log \left( M_{\rm star} [M_\odot] \right) = 8.83^{+0.22}_{-0.30}$,
$\log \left({\rm Age [yr]} \right) = 8.61^{+0.25}_{-0.35}$\footnote{The
age suggests that the formation redshift of this object is $\geq 8.5$.
},
and $E(B-V)_\star = 0.00^{+0.01}_{-0.00}$.
However, we find that the data are also well reproduced by a low-mass
($\log \left( M_{\rm star} [M_\odot] \right) = 7.72^{+0.07}_{-0.11}$),
extremely young
($\log \left({\rm Age [yr]} \right) = 6.16^{+0.30}_{-1.06}$),
and almost extinction free ($E(B-V)_\star = 0.04^{+0.02}_{-0.03}$)
model, which are also shown in Table \ref{tab:SEDfitting}.
Although these two models are extremely different, they have
almost the same $\chi^2$ values, and we cannot conclude which
is more favored.
In summary, the $z$-dropout galaxy is either a very young, low-mass
object, or a very old, massive object.
To be interesting, just like in the case of the $z=6.6$ LAEs,
models with intermediate ages and masses are not favored.

If we take account of the difference in the assumed star formation
history,
our results are broadly consistent with those of \cite{schaerer2010}.
As shown in Figure C.2 of their paper,
they have derived two equally-fitted solutions:
very young ($\sim 4$ Myr) and old ($\sim 700$ Myr).
Note that their $\chi^2$ values do not match with ours.
This would be because
they have parameterized the star-formation $e$-folding time
and metallicity, which we fix, as well,
and they have considered $B_{435}$, $V_{606}$, and $i_{775}$ data,
which we do not consider.

The top panel of Figure \ref{fig:Mstar_sSFR_age}
plots the best-fit age against the best-fit $M_{\rm star}$
for our LAEs and the $z \sim 7$ $z$-dropout.
The best-fit solutions for our $z=5.7$ and $6.6$ LAEs
(red filled circle and square)
suggest that they are low-mass and very young star-forming galaxies.
As for the $z \sim 7$ $z$-dropout,
both of the two extreme solutions (red and blue filled triangles)
are plotted.
If the younger solution is true,
the $z$-dropout is as old and massive as our LAEs.
On the other hand, if the older solution is true,
the $z$-dropout is much older and more massive than our LAEs.

The bottom panel of Figure \ref{fig:Mstar_sSFR_age}
shows the specific SFR (SSFR) as a function of $M_{\rm star}$.
Our $z=5.7$ and $6.6$ LAEs have low stellar masses
and very high SSFRs.
On the other hand, the $z$-dropout has either almost the same
$M_{\rm star}$ and SSFR as the LAEs, or far more massive and
lower SSFR than the LAEs.

\begin{figure}
 \begin{center}
  \includegraphics[scale=0.8]{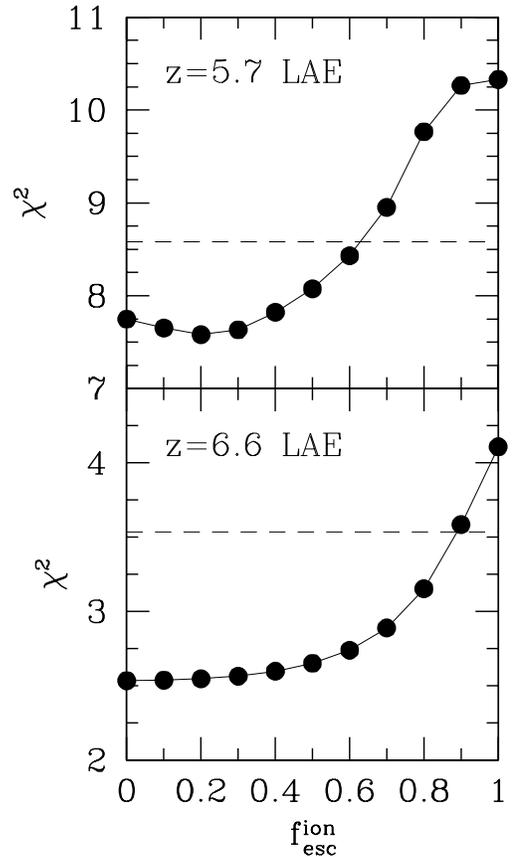}
\end{center}
 \caption[]
{
$\chi^2$ versus the Lyman-continuum escape fraction $f_{\rm esc}^{\rm ion}$
for the $z=5.7$ LAE (top panel) and the $z=6.6$ LAE (bottom panel).
The dashed lines correspond to $\chi^2_{\rm min} + 1$.
}
\label{fig:fesc_chi2}
\end{figure}

\begin{deluxetable*}{cccccccc} 
\tablecolumns{8} 
\tablewidth{0pt} 
\tablecaption{SED Fitting Results for the LAEs at $z=5.7$ and $6.6$ \label{tab:SEDfitting_fesc05}}
\tablehead{
    \colhead{model}    & \colhead{$Z$} & \colhead{$\log M_{\rm star}$}    & \colhead{$E(B-V)_\star$} & \colhead{$\log$(Age)}    
    & \colhead{$\log$(SFR)}    & \colhead{$\log$(SSFR)}  & \colhead{$\chi^2$} \\
    \colhead{}    & \colhead{[$Z_\odot$]}    & \colhead{$[M_\odot]$}    & \colhead{[mag]} & \colhead{[yr]}    
    & \colhead{[$M_\odot$ yr$^{-1}$]}  & \colhead{[yr$^{-1}$]}  & \colhead{}
}
\startdata 
\multicolumn{8}{c}{$z=5.7$ LAE} \\ 
\hline \\
$f_{\rm esc}^{\rm ion} = 0.2$
& $0.2$  & $7.55^{+0.15}_{-0.08}$ &  $0.00^{+0.04}_{-0.00}$  
& $5.95^{+0.91}_{-0.85}$  &  $1.60^{+0.96}_{-0.92}$  &  $-5.95^{+0.85}_{-2.24}$  &  $7.58$
\\
\hline \\
 \multicolumn{8}{c}{$z=6.6$ LAE} \\ 
 \hline \\ 
$f_{\rm esc}^{\rm ion} = 0.2$
& $0.2$  & $8.08^{+0.81}_{-0.28}$ &  $0.14^{+0.23}_{-0.11}$  
&  $5.95^{+1.41}_{-0.85}$  &  $2.13^{+1.23}_{-1.04}$  &  $-5.95^{+0.85}_{-1.38}$  &  $2.55$  
\enddata 
\end{deluxetable*} 

\begin{figure*}
 \begin{center}
  \includegraphics[scale=1.0]{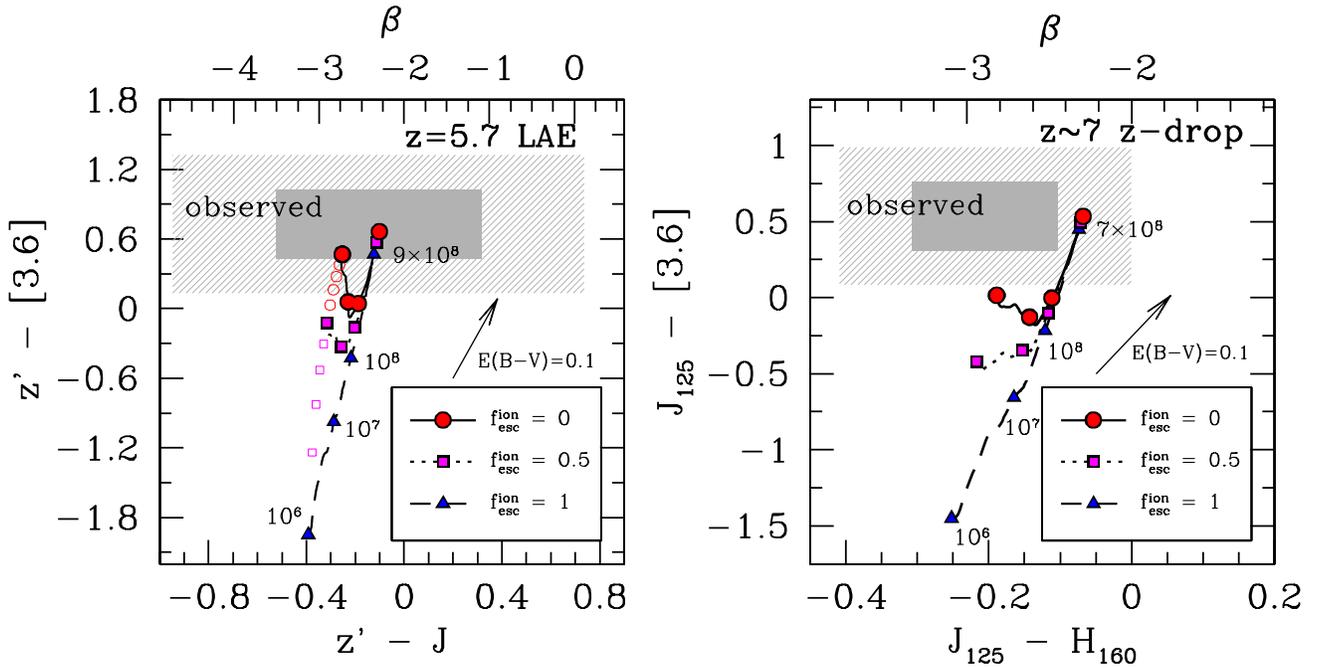}
\end{center}
 \caption[]
{
\textbf{Left panel}:
comparison of the observed $z'-[3.6]$ versus $z'-J$ of the $z=5.7$ LAE
(shaded region: $1\sigma$ range; hatched region: $2\sigma$ range)
with model tracks from $\simeq 10^6$ yr to $9 \times 10^8$ yr
with $Z=0.2 Z_\odot$ and a constant star formation rate,
for $f_{\rm esc}^{\rm ion} = 0$ (solid curve with red circles corresponding
to four ages: $10^6$, $10^7$, $10^8$, $9 \times 10^8$ yr
from left to right),
$0.5$ (dotted curve with magenta squares),
and $1$ (dashed curve with blue triangles).
The open symbols represent $\simeq 10^6$ yr models
with $f_{\rm esc}^{\rm ion} = 0.1$ -- $0.4$ (red open circles)
and $f_{\rm esc}^{\rm ion} = 0.6$ -- $0.9$ (magenta open squares)
from top to bottom.
The arrow shows the effects of dust extinction on model colors
\citep{calzetti2000}.
\textbf{Right panel}:
comparison of the observed $J_{125}-[3.6]$ versus $J_{125} - H_{160}$
of the $z\sim 7$ $z$-dropout \citep{labbe2010}
(shaded region: $1\sigma$ range; hatched region: $2\sigma$ range)
with model tracks from $\simeq 10^6$ yr to $7 \times 10^8$ yr
with $Z=0.2 Z_\odot$ and a constant star formation rate,
for $f_{\rm esc}^{\rm ion} = 0$ (solid curve with red circles corresponding
to four ages: $\simeq 10^6$, $10^7$, $10^8$, $7 \times 10^8$ yr
from left to right),
$0.5$ (dotted curve with magenta squares),
and $1$ (dashed curve with blue triangles).
The arrow shows the effects of dust extinction on model colors
\citep{calzetti2000}.
}
\label{fig:fesc_color_color}
\end{figure*}

\subsection{Constraints on Ly Continuum Escape Fraction}
\label{subsec:Implication_for_fesc}

In the previous sections we consider two extreme values for
the Ly continuum escape fraction:
$f_{\rm esc}^{\rm ion} = 1$ (pure stellar)
and $f_{\rm esc}^{\rm ion} = 0$ (stellar $+$ nebular).
Indeed, our data are not deep enough to place as a strong
constraint on $f_{\rm esc}^{\rm ion}$ as on the other parameters.
However, since $f_{\rm esc}^{\rm ion}$ is a very important quantity which
controls cosmic reionization, in this subsection we perform SED
fitting with $f_{\rm esc}^{\rm ion}$ as an additional free parameter,
to try to obtain rough constraints on $f_{\rm esc}^{\rm ion}$.
We vary $f_{\rm esc}^{\rm ion}$ over 0 and 1 with an interval of 0.1.
In the fitting of the $z=6.6$ LAE, we search for the best-fit age
in the range of $< 20$ Myr following the argument in Section \ref{subsec:stellarpop_of_LAEs}
that the $z=6.6$ LAE is likely to be very young.

Figure \ref{fig:fesc_chi2} shows $\chi^2$ as a function of $f_{\rm esc}^{\rm ion}$
for our LAEs.
For $z=5.7$, $\chi^2$ is nearly constant up to $f_{\rm esc}^{\rm ion}=0.5$
and then starts to increase.
Beyond $f_{\rm esc}^{\rm ion} \sim 0.6$ it exceeds $\chi^2_{\rm min} + 1$,
the $1\sigma$ confidence level.
We can thus place an upper limit of $f_{\rm esc}^{\rm ion} \sim 0.6$.
For $z=6.6$, $\chi^2$ changes little up to as high as
$f_{\rm esc}^{\rm ion}= 0.8$, and exceeds $\chi^2_{\rm min} + 1$ at
around 0.9.
Thus $f_{\rm esc}^{\rm ion}$ is only loosely constrained as
$f_{\rm esc}^{\rm ion} \lesssim 0.9$.
We also do a similar analysis to the $z$-dropout but obtain no
meaningful constraint, since $\chi^2$ does not change larger than
unity over the whole $f_{\rm esc}^{\rm ion}$ range.
This is probably because the $z$-dropout is fit well by an old-age
model as well (see Section \ref{subsec:stellarpop_of_LBGs}) in which
nebular emission is so weak that $\chi^2$ is insensitive to
$f_{\rm esc}^{\rm ion}$.
If we limit the age of the $z$-dropout to $<20$ Myr, as in the
case of the $z=6.6$ LAE, then we obtain an upper limit of
$f_{\rm esc}^{\rm ion} \sim 0.2$.

Although our LAE data cannot strongly constrain $f_{\rm esc}^{\rm ion}$, 
it would be worth presenting the best-fit results for $f_{\rm esc}^{\rm ion} = 0.2$, 
at which $\chi^2$ for the $z=5.7$ LAE reaches its minimum. 
As found in Table \ref{tab:SEDfitting_fesc05}, 
the best-fit parameters for $f_{\rm esc}^{\rm ion}=0.2$
overlap well with those for $f_{\rm esc}^{\rm ion}=0$
(Table \ref{tab:SEDfitting}) within the errors, 
indicating that our conclusions on the stellar populations 
obtained in Section 5.1 are relatively robust against 
the uncertainty in $f_{\rm esc}^{\rm ion}$.

The reason for $f_{\rm esc}^{\rm ion}$ having an upper limit is that
with too large $f_{\rm esc}^{\rm ion}$, young stellar populations
favored by the observed blue UV continua cannot produce
nebular emission strong enough to account for the observed
red UV-to-optical color.
In this sense, the UV-to-optical color is critical
to constrain $f_{\rm esc}^{\rm ion}$.
We explain this situation using Figure \ref{fig:fesc_color_color}.
In this figure, three model tracks for the $z=5.7$ LAE over ages
of $10^6$ to $9 \times 10^8$ yr are plotted according to
three $f_{\rm esc}^{\rm ion}$ values, $0, 0.5, 1$,
in the $z'-[3.6]$ versus $z'-J$ plane.
We take $z'-J$ as a representative of the rest-frame UV color
and $z'-[3.6]$ as a rest-frame UV-to-optical color bracketing 4000 {\AA}.
The shaded and hatched regions are, respectively,
the observationally permitted $1\sigma$ and $2\sigma$ ranges.
The $z'-J$ color at a fixed age becomes bluer with $f_{\rm esc}^{\rm ion}$
but the change is modest.
In contrast, $z'-[3.6]$ increases with $f_{\rm esc}^{\rm ion}$ very
sensitively for young ages.
For example, for an age of $10^6$ yr, which is close to the
best-fit age of the $z=5.7$ LAE, $z'-[3.6]$ becomes rapidly
redder with $f_{\rm esc}^{\rm ion}$ (magenta squares) and the model
goes out of the hatched region at $f_{\rm esc}^{\rm ion} \sim 0.3$.
Although the goodness of models should be measured using the
all magnitude data, this figure demonstrates the importance
of a UV-to-optical color in constraining $f_{\rm esc}^{\rm ion}$.

Recently, \cite{bouwens2010b} proposed that the very blue UV color
of $z$-dropout galaxies they found may be due to weak nebular
emission and hence high $f_{\rm esc}^{\rm ion}$, because strong nebular
emission makes the UV color too red.
However, as they already state, $f_{\rm esc}^{\rm ion}$ is not uniquely
determined from the UV color but it also depends on the age of
the stellar population.
Inclusion of a UV-to-optical color greatly reduces this
age-$f_{\rm esc}^{\rm ion}$ degeneracy, as illustrated in
Figure \ref{fig:fesc_color_color}.
Basically, with very accurate measurements of a UV color
and a UV-to-optical color, we can obtain a stringent constraint
on $f_{\rm esc}^{\rm ion}$ (for a fixed metallicity and IMF).
For instance, if $\beta$ of the $z$-dropout was found to be
$-3$ with a great accuracy, low $f_{\rm esc}^{\rm ion}$ models 
such as $f_{\rm esc}^{\rm ion}=0$ would be ruled out. 
On the other hand, if $\beta$ and
$J_{125} - [3.6]$ of the $z$-dropout were found to be $-2.8$
and $0$ respectively, with great accuracies, low $f_{\rm esc}^{\rm ion}$
models would be favored.

Several studies have recently constrained lower limits of the escape
fraction based on the UV luminosity density of $z$-dropout galaxies;
\cite{ouchi2009b} have obtained $f_{\rm esc}^{\rm ion} \gtrsim 0.2$,
and \cite{finkelstein2009f} have obtained $f_{\rm esc}^{\rm ion} \gtrsim 0.3$.
The upper limits obtained above complement these lower limits,
thus narrowing the range permitted.

Several theoretical studies have argued that
$f_{\rm esc}^{\rm ion}$ could vary depending on their host dark halo masses.
However, there is no consensus on this tendency;
\cite{wise2009} have reported that
$f_{\rm esc}^{\rm ion}$ decreases as the halo mass decreases,
while \cite{razoumov2010} have shown
an opposite tendency \citep[see also,][]{yajima2010}.
This controversy may be resolved if our method is applied to
objects in a wide mass range.

\subsection{Contribution of LAEs to the Stellar Mass Density and the Cosmic Star Formation Rate Density}

We estimate the contribution from LAEs to the cosmic stellar mass
density by dividing the total stellar mass of LAEs
by the comoving volume searched by each narrow band.
The total stellar mass at each redshift is defined as
the stellar mass of the stacked LAEs multiplied by their number.
Assuming $f_{\rm esc}^{\rm ion}=0$,
we obtain $\simeq 8.5 \times 10^3$ [$M_\odot$ Mpc$^{-3}$] for
$z=5.7$ LAEs and
$\simeq 1.6 \times 10^4$ [$M_\odot$ Mpc$^{-3}$] for $z=6.6$ LAEs.
These values should be taken as lower limits, since we have
excluded objects detected and/or significantly confused by
neighboring objects in the $3.6\mu$m image.

\cite{eyles2007} have summed up the stellar masses of the $i$-drop
galaxies they detected, to obtain a lower limit to the stellar mass
density of $z \sim 6$ galaxies of $\sim 3 \times 10^6$ [$M_\odot$ Mpc$^{-3}$]. 
\cite{stark2009} have derived the mass function of $i'$-dropout galaxies 
and obtained a stellar mass density of 
$\simeq 4.9 \times 10^6$  [$M_\odot$ Mpc$^{-3}$]
by integrating the mass function brightward of $M_{1500} = -20$.
The stellar mass density of our $z=5.7$ LAEs is only 
$\sim 0.2 - 0.3${\%}
of 
these values.

For $z \sim 7$ galaxies,
\cite{labbe2010} have obtained a lower limit of
$\sim 3.7 \times 10^6$ [$M_\odot$ Mpc$^{-3}$]
by multiplying the UV luminosity density of $z$-dropout galaxies
given in \cite{bouwens2010} by an estimated mass-to-luminosity ratio.
Again, the stellar mass density of our $z=6.6$ LAEs is as low as
$\sim 0.4${\%} of this value.

Using SED fitting similar to ours, 
\cite{ono2009} have obtained 
$1.4 \times 10^5$ [$M_\odot$ Mpc$^{-3}$] for $z=3.1$ LAEs, 
and 
$5.2 \times 10^5$ [$M_\odot$ Mpc$^{-3}$] for $z=3.7$ LAEs.
These values are about $10 - 60$ times larger 
than those of $z=5.7$ and $6.6$ LAEs, 
although they assumed $f_{\rm esc}^{\rm ion}=1$.

Similarly, we estimate the contribution from LAEs to the cosmic
star formation rate density (SFRD), where the total star formation rate
is computed as the star formation rate of the stacked objects
multiplied by their number.
We obtain $\simeq 2.7 \times 10^{-3}$ [$M_\odot$ yr$^{-1}$ Mpc$^{-3}$]
for $z=5.7$ LAEs
and $\simeq 1.8 \times 10^{-2}$ [$M_\odot$ yr$^{-1}$ Mpc$^{-3}$]
for $z=6.6$ LAEs.
These values are also lower limits to the total stellar mass density of LAEs.

\cite{bouwens2007} have estimated the SFRD of $z \sim 6$ galaxies
to be $\sim 7.2 \times 10^{-3}$ [$M_\odot$ yr$^{-1}$ Mpc$^{-3}$]
by integrating the UV luminosity function of $i$-dropout galaxies.
Similarly, \cite{ouchi2009b} have obtained
$\sim 7.5 \times 10^{-3}$ [$M_\odot$ yr$^{-1}$ Mpc$^{-3}$] for
$z \sim 7$ galaxies from the UV luminosity function of $z$-dropout galaxies \citep[see also,][]{bouwens2010}.
These two values are comparable to the lower limits obtained for
our LAEs at $z=5.7$ and $6.6$, suggesting that LAEs are major
sources of the cosmic star formation at $z \sim 6$ -- $7$.
\cite{ono2009} have obtained 
$\simeq 8.2 \times 10^{-3}$ [$M_\odot$ yr$^{-1}$ Mpc$^{-3}$] for $z=3.1$ LAEs, 
and 
$\simeq 1.3 \times 10^{-1}$ [$M_\odot$ yr$^{-1}$ Mpc$^{-3}$] for $z=3.7$ LAEs.
These values are about $0.5 - 50$ times larger 
than those of $z=5.7$ and $6.6$ LAEs.

\section{CONCLUSIONS} \label{sec:conclusion}

In this paper, we investigated the stellar populations of LAEs at
$z = 5.7$ and $6.6$
found in $0.65$ deg$^2$ of the SXDF,
based on deep rest-frame UV-to-optical photometry
obtained from the three surveys: SXDS, UKIDSS/UDS, and SpUDS.
We made composite images from $165$ and $91$ LAEs at $z=5.7$ and
$6.6$, respectively,
which are fainter than the $3\sigma$ magnitude in the IRAC
$3.6\mu$m band,
and derived typical SEDs of $z\sim 6-7$ LAEs for the first time.
We found that their UV continua are as blue as those of dropout
galaxies at similar redshifts, with UV spectral slopes $\beta\sim -3$,
albeit with large photometric uncertainties.
Fitting stellar population synthesis models with and without
nebular emission, which is parameterized by $f_{\rm esc}^{\rm ion}$,
to the multiband data of the stacked objects at $z=5.7$ and $6.6$,
we derived their stellar masses, ages, and dust extinction.

Our main results are as follows:

\begin{enumerate}
\item[(i)]
We find that the stacked LAEs at both redshifts are fitted well
by $f_{\rm esc}^{\rm ion}=0$ models.
The best-fit $f_{\rm esc}^{\rm ion}=0$ models have low stellar masses of
$\sim \left( 3-10 \right) \times 10^{7}M_\odot$, very young ages of $\sim 1-3$ Myr,
and negligible dust extinction.
In these models, young stellar populations reproduce the observed
blue UV continua, and strong nebular emission redward of 4000 {\AA} 
makes the UV-to-optical color as red as observed.

While we find that the $z=6.6$ LAE is also fitted similarly well
by an old, massive model without nebular emission, we do not take
this models as the best-fit model, since typical LAEs up to
$z \sim 6$, including our $z=5.7$ LAEs, have been consistently
found to be very young and low-mass galaxies.
We propose that typical $z\sim 6-7$ LAEs are candidates of
galaxy building blocks at the early stage of galaxy formation.

\item[(ii)] We estimate the Ly$\alpha$ escape fraction to be
$f_{\rm esc}^{{\rm Ly}\alpha} \simeq 0.36$ for the $z=5.7$ LAE
and $f_{\rm esc}^{{\rm Ly}\alpha} \simeq 0.04$ for the $z=6.6$ LAE
with large errors.
This decrease from $z=5.7$ to $6.6$, if real, might be due to
an increase in the neutral fraction of the IGM.

\item[(iii)] We also apply SED fitting
to a stacked object from $z \sim 7$ $z$-dropout galaxy candidates
found by a recent HST/WFC3 deep survey,
and find that it is either a young, low-mass galaxy with
strong nebular emission similar to the $z=5.7$ and 6.6 LAEs,
or a very old and massive galaxy with little nebular emission.

\item[(iv)] From the constraints on nebular emission models,
we estimate the upper limit of the Ly-continuum escape fraction
to be $f_{\rm esc}^{\rm ion} \sim 0.6$ for the $z=5.7$ LAE
and $\sim 0.9$ for the $z=6.6$ LAE.
We apply this technique to the $z$-dropout galaxies,
but obtain no meaningful constraints on $f_{\rm esc}^{\rm ion}$.
\end{enumerate}

\section*{Acknowledgements}

We would like to thank the anonymous referee for helpful comments and suggestions. 
We also thank Jason Kalirai for kindly providing the data; 
Daniel Schaerer, Richard Ellis, Andrea Ferrara, Eric Gawiser, and Daniel Stark  for useful conversation.
M.O. has been supported via Carnegie Fellowship.
JSD and RJM thank the Royal Society for support via a Wolfson Research Merit
Award and a University Research Fellowship respectively.


\bibliographystyle{apj}
\bibliography{apj-jour,../../../Papers/papers_gal/ms}

\end{document}